%% file: OrdinalRandomForests2.tex
\documentclass[12pt,indent]{article}%a5paper
\usepackage{a4,latexsym,amsmath}%a4
\usepackage[latin1]{inputenc}
\usepackage{float}
\usepackage{natbib}
\usepackage{graphics}
\usepackage[english]{babel}
\usepackage{tabularx}
\usepackage{lscape}
\usepackage{multirow}
\usepackage{rotating}
\usepackage{dsfont}
\usepackage{subfig}
\usepackage{bbm}
\usepackage[table]{xcolor}

\usepackage{amsmath}
\usepackage{amsfonts}

\usepackage{calc}
\usepackage{ifthen}

\usepackage{%
  longtable,
  array,
  booktabs,
  dcolumn,
  rotating,
  shortvrb,
  tabularx,
  units,
  multirow
}

\newenvironment{tabularsmall}
{ \footnotesize \sffamily \tabular } {
\endtabular
\normalfont }
\input{def}

\usepackage{natbib}

\usepackage{setspace}
\begin{document}
\bibliographystyle{chicago}
\sloppy
%%%%%%%%%%%%%%%%%%%%%%%%%%%%%%%%%%%%%%%%%%%%%%%%%%%%%%%%%%%%%%%%%%%%%
%                                                                   %
%         Definition einer modifizierten Kapitelï¿½berschrift         %
%                                                                   %

\makeatletter
\renewcommand{\section}{\@startsection{section}{1}{\z@}%
        {-3.5ex \@plus -1ex \@minus -.2ex}%
        {1.5ex \@plus.2ex}%
        {\reset@font\Large\sffamily}}
\renewcommand{\subsection}{\@startsection{subsection}{1}{\z@}%
        {-3.25ex \@plus -1ex \@minus -.2ex}%
        {1.1ex \@plus.2ex}%
        {\reset@font\large\sffamily\flushleft}}
\renewcommand{\subsubsection}{\@startsection{subsubsection}{1}{\z@}%
        {-3.25ex \@plus -1ex \@minus -.2ex}%
        {1.1ex \@plus.2ex}%
        {\reset@font\normalsize\sffamily\flushleft}}
\makeatother

%                                                                   %
%%%%%%%%%%%%%%%%%%%%%%%%%%%%%%%%%%%%%%%%%%%%%%%%%%%%%%%%%%%%%%%%%%%%%

%%%%%%%%%%%%%%%%%%%%%%%%%%%%%%%%%%%%%%%%%%%%%%%%%%%%%%%%%%%%%%%%%%%%%
%                                                                   %
%         Definition einer modifizierten Bildunterschrift           %
%                                                                   %

\newsavebox{\tempbox}
\newlength{\linelength}
\setlength{\linelength}{\linewidth-10mm} \makeatletter
\renewcommand{\@makecaption}[2]
{
  \renewcommand{\baselinestretch}{1.1} \normalsize\small
  \vspace{5mm}
  \sbox{\tempbox}{#1: #2}
  \ifthenelse{\lengthtest{\wd\tempbox>\linelength}}
  {\noindent\hspace*{4mm}\parbox{\linewidth-10mm}{\sc#1: \sl#2\par}}
  {\begin{center}\sc#1: \sl#2\par\end{center}}
}

%                                                                   %
%%%%%%%%%%%%%%%%%%%%%%%%%%%%%%%%%%%%%%%%%%%%%%%%%%%%%%%%%%%%%%%%%%%%%

%\bibliographystyle{chicago}
%\baselineskip7mm
%\parindent 0.5cm
%\parskip2ex plus0.5ex minus 0.5ex
%\setlength{\parskip}{7pt plus 1pt minus 1pt}

\def\R{\mathchoice{ \hbox{${\rm I}\!{\rm R}$} }
                   { \hbox{${\rm I}\!{\rm R}$} }
                   { \hbox{$ \scriptstyle  {\rm I}\!{\rm R}$} }
                   { \hbox{$ \scriptscriptstyle  {\rm I}\!{\rm R}$} }  }

\def\N{\mathchoice{ \hbox{${\rm I}\!{\rm N}$} }
                   { \hbox{${\rm I}\!{\rm N}$} }
                   { \hbox{$ \scriptstyle  {\rm I}\!{\rm N}$} }
                   { \hbox{$ \scriptscriptstyle  {\rm I}\!{\rm N}$} }  }

\def\d{\displaystyle}

\title{Ordinal Trees and Random Forests:  Score-Free Recursive Partitioning and Improved Ensembles}
\author{Gerhard Tutz \\{\small Ludwig-Maximilians-Universit\"{a}t M\"{u}nchen}\\
 \small Akademiestra{\ss}e 1, 80799 M\"{u}nchen }

%\author{Jan Gertheiss\footnote{To whom correspondence should be
%addressed: \texttt{jan.gertheiss@stat.uni-muenchen.de.}}
%\footnote{Department of Statistics, Ludwig-Maximilians-Universität
%Munich, Germany.} \ \& Gerhard Tutz\footnotemark[2]}

% \ead{tutz@stat.uni-muenchen.de}
% \address{Ludwig-Maximilians-Universit\"{a}t M\"{u}nchen, Ludwigstra{\ss}e 33, D-80539 M\"{u}nchen, Germany}
%\author{Lorenz Uhlmann}
%\ead{tutz@stat.uni-muenchen.de}
%\address[muc]{Ludwig-Maximilians-Universit\"{a}t M\"{u}nchen, Akademiestra{\ss}e 1, 80799 M\"{u}nchen, Germany}
%\cortext[cor]{Corresponding author. Tel.: ++4989 2180 3044; fax.:
%++4989 2180 5308.}
%{\texttt{\small \{tutz, uhlmann\}@stat.uni-muenchen.de}}}
%\address[muc1]{Ludwig-Maximilians-University Munich, Ludwigstrasse 33, D-80539 Munich, Germany}
%\address[muc2]{Ludwig-Maximilians-University Munich, Akademiestra{\ss}e 1, D-80799 Munich, Germany}
%\cortext[cor]{Corresponding author. Tel.: ++49 89 2180 3044; fax.:
%++49 89 2180 ???.}
\maketitle

\begin{abstract} % \renewcommand{\baselinestretch}{1.3} \small\normalsize
\noindent
%\begin{center}
Existing ordinal trees and random forests typically use scores that are assigned to the ordered categories, which implies that a higher scale level is used.
Versions of ordinal trees are proposed that take the scale level seriously and avoid the assignment of artificial scores. The basic construction principle is based on an investigation of the binary models that are implicitly used in parametric ordinal regression. These building blocks can be fitted by trees and combined in a similar way as in parametric models. The obtained trees use the ordinal scale level only. Since binary trees and random forests are constituent elements of the trees one can exploit the wide range of binary trees that have already been developed. A further topic is the potentially poor performance of random forests, which seems to have ignored in the literature. Ensembles that include parametric models are proposed to obtain prediction methods that tend to perform well in a wide range of settings. The performance of the methods is evaluated empirically by using   several data sets.

\end{abstract}


\section{Introduction}

There is a long tradition of analyzing ordinal response data by using parametric models, which started  with the seminal paper of \citep{McCullagh:80}.
More recently, recursive partitioning methods have been developed that allow to investigate the impact of explanatory variables by non-parametric tools.
Single and random trees for ordinal responses have several advantages, they can be applied to large data sets and are considered to perform very well in prediction.

A problem with most of the ordinal trees is that they assume that scores are assigned to the ordered categories of the response. The assignment of scores can be warranted in some cases, in particular if ordinal responses are built from continuous variables by grouping. However, it is rather artificial and arbitrary in genuine ordinal response data, for example, if the response represents ordered levels of severeness of a disease. Then one can not choose the midpoints of the intervals from which the ordered response is built as suggested by \citet{Hotetal:2006} since no continuous variable is observed.  If nevertheless scores are assigned they can affect the prediction results although that has not always to be the case, see also \citet{janitza2016random}. 

The packages \textit{rpartOrdinal} \citep{archer2010rpartordinal} as well as the improved version \textit{ rpartScore}  \citep{galimberti2012classification}, which are based on the Gini impurity function,  use assigned scores. The same holds for the random forests proposed by \citet{janitza2016random} and the ordinal version of conditional trees of the package \textit{party} \citep{Hotetal:2006,hothorn2015partykit}.  The random forest approach proposed by \citet{hornung2019ordinal} is somewhat different, it also translates ordinal measurements into continuous scores but optimizes scores instead of using a fixed score. Versions of random forests without scores were proposed more recently by \citet{buri2020model}. They use the ordinal proportional odds model to obtain statistics that are used in splitting.

In the following  alternative trees and random forests that take the scale level of the response seriously are proposed. The main concept is that ordinal responses  contain binary responses as building blocks.  This has already been implicitly used in parametric modeling approaches. For example, the widely used proportional odds model can be seen as a model that parameterizes the split of response categories into two groups of adjacent categories. But the principle also holds for alternative models as the adjacent categories model and the sequential model, see \citet{TuHetWire21} for an overview and a taxonomy of ordinal regression models.  The proposed trees explicitly use the representation of ordinal responses as a set of binary variables. Random forests for the binary variables are used to obtain random forests for ordinal response data.

%see \citet{TutzBook2011}
For random forests it is important that they provide good performance in terms of prediction. They are commonly considered as being very efficient. However, as will be demonstrated this does not hold in general. In many cases simple parametric models turn out to be at least as efficient and sometimes more efficient than the carefully designed random forests.  Typically, when ordinal forests are propagated the accuracy is investigated for versions of random forests only but they are not compared to parametric competitors. In the following we propagate the use of ensembles that include parametric models to provide a stable prediction tool that works well in all kinds of data sets.

The paper has two objectives, introducing score-free recursive partitioning and random forests, and propagating ensembles that include parametric models. In Section \label{sec:binrep} the representation of ordinal responses as a sequence of binary responses is briefly considered. It makes clear that specific binary responses  can be seen as building blocks of classical parametric models. In Section \ref{sec:rec} it is shown how these building blocks can be used to construct score-free trees and random forests. In addition, more general ensembles are considered. In Section \ref{sec:pred} the performance of the ensembles is investigated by using real data sets. Section \ref{sec:imp} is devoted to importance measures, which are an essential ingredient of random forests since the impact of variables on prediction in random forests is not directly available.

%\citet{galimberti2012package},

\section{Binary Representations of  Ordinal Responses}\label{sec:binrep}

In the following the representation of ordinal response as a collection of binary responses is considered. It can be seen as being behind the construction of parametric ordinal models and will serve to construct a novel type of recursive partitioning that does not use assigned scores.

Let the ordinal response $Y$ take  values from $\{1,\dots,k\}$. Although these values suggest a univariate response the actual response is multivariate since the numbers $1,\dots,k$ just represent that  outcomes are ordered but  distances between numbers assigned to categories should not be built  in an ordinal scale because they are not interpretable.

%\subsection{Representations by Split Variables}
A multivariate representation of the outcome can be obtained by using binary dummy variables. Natural candidates for dummy variables are the split variables
\begin{equation}\label{eq:cond2}
Y_{r}=\left\{
\begin{array}{rr}
1&Y_{ }  \ge r \\
0&Y_{ } < r.
\end{array} 
\right.
\end{equation}
%\[Y_r=1 \text{ if }  Y \ge r,  Y_r=0 \text{ otherwise }.\]
Then $Y=r$ is represented by a sequence of ones   followed by a sequence of zeros,
\[
(Y_{2} \dots,Y_{k})=(1,\dots,1,0,\dots,0).
\]
The vector $(Y_{2} \dots,Y_{k})$ can be seen as a multivariate representation of the response. The dummy variables that generate vectors of this form, which are characterized by a sequence of ones   followed by a sequence of zeros have also be referred to as Guttman variables \citep{andrich2013expanded}.

Classical ordinal regression model  use these dummy variables but are most often derived from the assumption of an underlying continuous variable, and the link to split variables is ignored. The most widely used \textit{proportional odds model}, also called \textit{cumulative logistic model},  has the form 
\begin{equation}\label{eq:cumpart}
P(Y \ge r |\xb)= F(\beta_{0r}+\xb^T\betab),   \quad r=2,\dots,k.
\end{equation}
where $\xb$ is a vector of explanatory variables and $F(\eta)= \exp(\eta)/(1+\exp(\eta))$ is the logistic distribution function. For the parameters one has the restriction $\beta_{02} \ge \dots \ge \beta_{0k}$. The model explicitly uses the dichotomizations given by (\ref{eq:cond2}). Since $Y \ge r$ iff $Y_r = 1$ the model can also be given as 
\begin{equation}\label{eq:bin}
P(Y_r=1|\xb )= F(\beta_{0r}+\xb^T\betab),  \quad r=2,\dots,k.
\end{equation}
Thus, the proportional odds model is equivalent to a collection of binary logit models that have to hold simultaneously. The model   implies that 
the effect of covariates contained in $\xb^T\betab$ is the same for all dichotomizations. That means if one fits the binary models (\ref{eq:bin}) separately one should obtain similar values for estimates of $\betab$. This restriction can be weakened by using the partial proportional odds model, in which the effect of variables may depend on the category, that is, the  linear term $\xb^T\betab$ in (\ref{eq:cumpart}) is replaced by $\xb^T\betab_r$.

Model (\ref{eq:cumpart}) is a so-called cumulative model since on the left hand side one has the sum of probabilities $P(Y \ge r |\xb)$. Cumulative models form a whole family of models, whose members are characterized by the choice of a specific  strictly increasing distribution function $F(.)$. They have been investigated and extended, among others,  by \citet{McCullagh:80}, \citet{Brant:90}, \citet{PetHar:90},   \citet{BenGro:98}, \citet{Cox:95}, \citet{kim2003assessing} and \citet{liu2009graphical}.

An alternative ordinal regression model is the \textit{adjacent categories model}, which has the basic form
\begin{equation}\label{eq:adj}
P(Y \ge r|Y \in \{ r-1,r\}, \xb)= F(\beta_{0r}+\xb^T\betab ), \quad r=2,\dots,k.
\end{equation}
Since $P(Y \ge r|Y \in \{ r-1,r\}, \xb)=P(Y = r|Y \in \{ r-1,r\}, \xb)$ it specifies the probability of observing category $r$ given the response is in categories $\{ r-1,r\}$.  Because of the conditioning it can be seen as a local model. The interesting point is that it also uses the split variabales. It is easily seen that it is equivalent to  
\begin{equation}\label{eq:adj3}
P(Y_r =1|Y_{r-1}=1, Y_{r+1}=0,\xb)= F(\beta_{0r}+\xb^T\betab ), \quad r=2,\dots,k.
\end{equation} 
Thus, it specifies the binary response variable $Y_r$ \textit{conditionally} in contrast to   cumulative models, which determine the binary response directly in an unconditional way. But as for cumulative models it is assumed that the binary models (\ref{eq:adj3}) hold simultaneously. 

The adjacent categories logit model may also be considered as the  regression model that is obtained from the row-column
(RC) association model considered by \citet{Goodman:81a,Goodman:81b}, \citet{kateri2014contingency}. It is also related to Anderson's stereotype model \citep{Anderson:84},  which was considered by \citet{Greenland:94}  and \citet{fernandez2019method}. It has been most widely used as a latent trait model in the form
of the partial credit model \citep{Masters:82, MasWri:84,muraki1997generalized}.

An advantage of the adjacent categories model is that one can replace the parameter vector $\betab$ by a category-specific parameter vector $\betab_r$ without running into problems. In cumulative models one has the restriction $P(Y \ge 2| \xb) \ge\dots P(Y \ge k| \xb)$, which can yield problems, in partivullar when fitting the binary models (\ref{eq:bin}) with   category-specific parameter vectors $\betab_r$.
For overviews of parametric ordinal models,  see, for example,  \citet{Agresti:2009}, \citet{TutzBook2011}. They also include a third type of ordinal model, the sequential model, which is a specific process model, which could also be extended to tree type models. But because of its specific nature we do not consider it explicitly.

The main point is that binary models are at the core of parametric ordinal models. There is a good reason for that because the splits represent the order in categories without assuming more than an order of categories. In the following this is exploited to construct trees that account for the ordering of categories.

\section{Recursive Partitioning  Based on Splits}\label{sec:rec}

The crucial role of split variables in modeling ordered response can be used to obtain non-parametric tree models that use the ordering efficiently. There are basically two ways to do so, one is by using the split variables directly, which corresponds to cumulative type models, the other approach is to use them conditionally, which corresponds to the adjacent categories approach.

\subsection{Trees for Split Variables}

\begin{figure}[h!]
\centering
\includegraphics[width=0.55\textwidth]{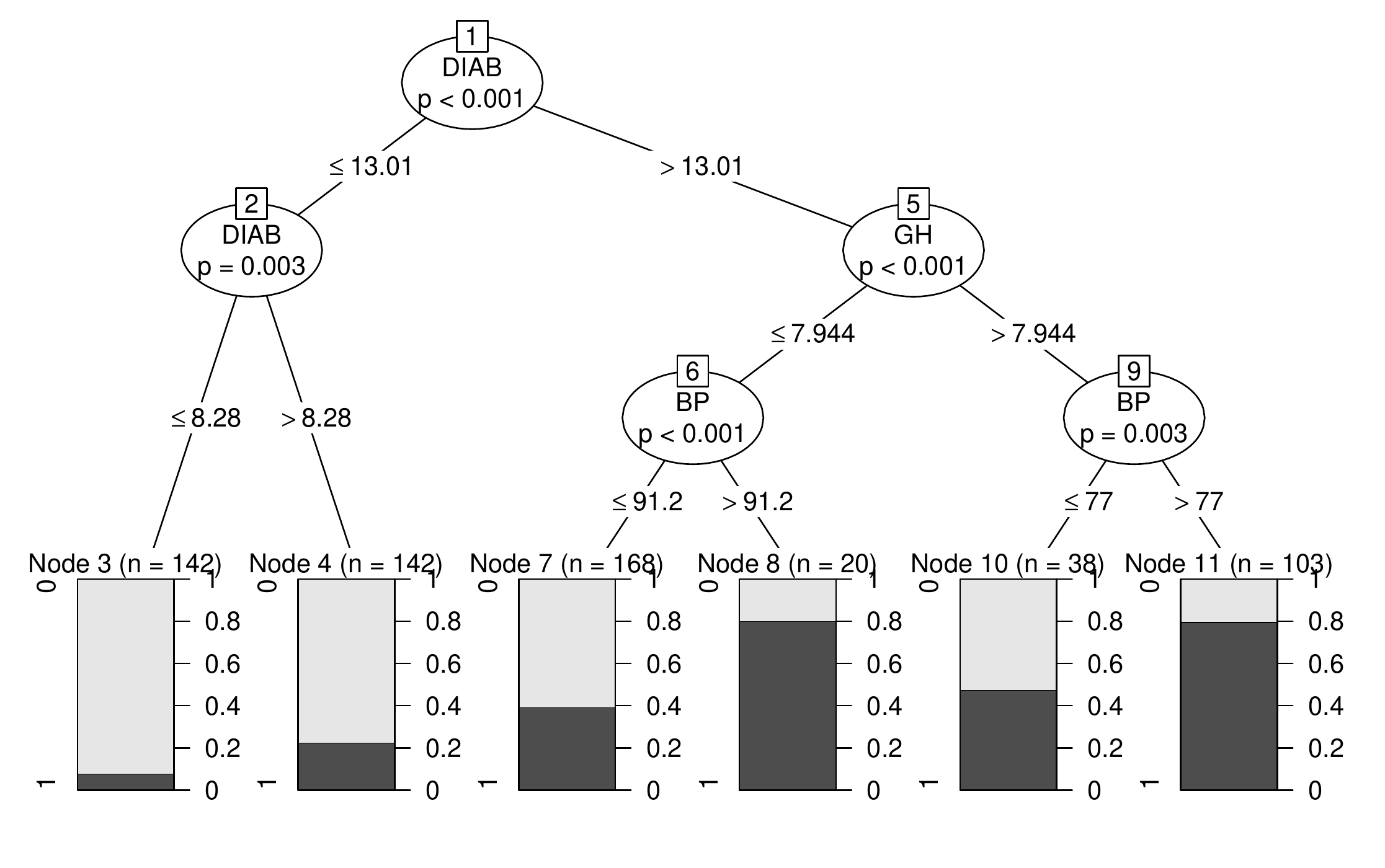}
\includegraphics[width=0.55\textwidth]{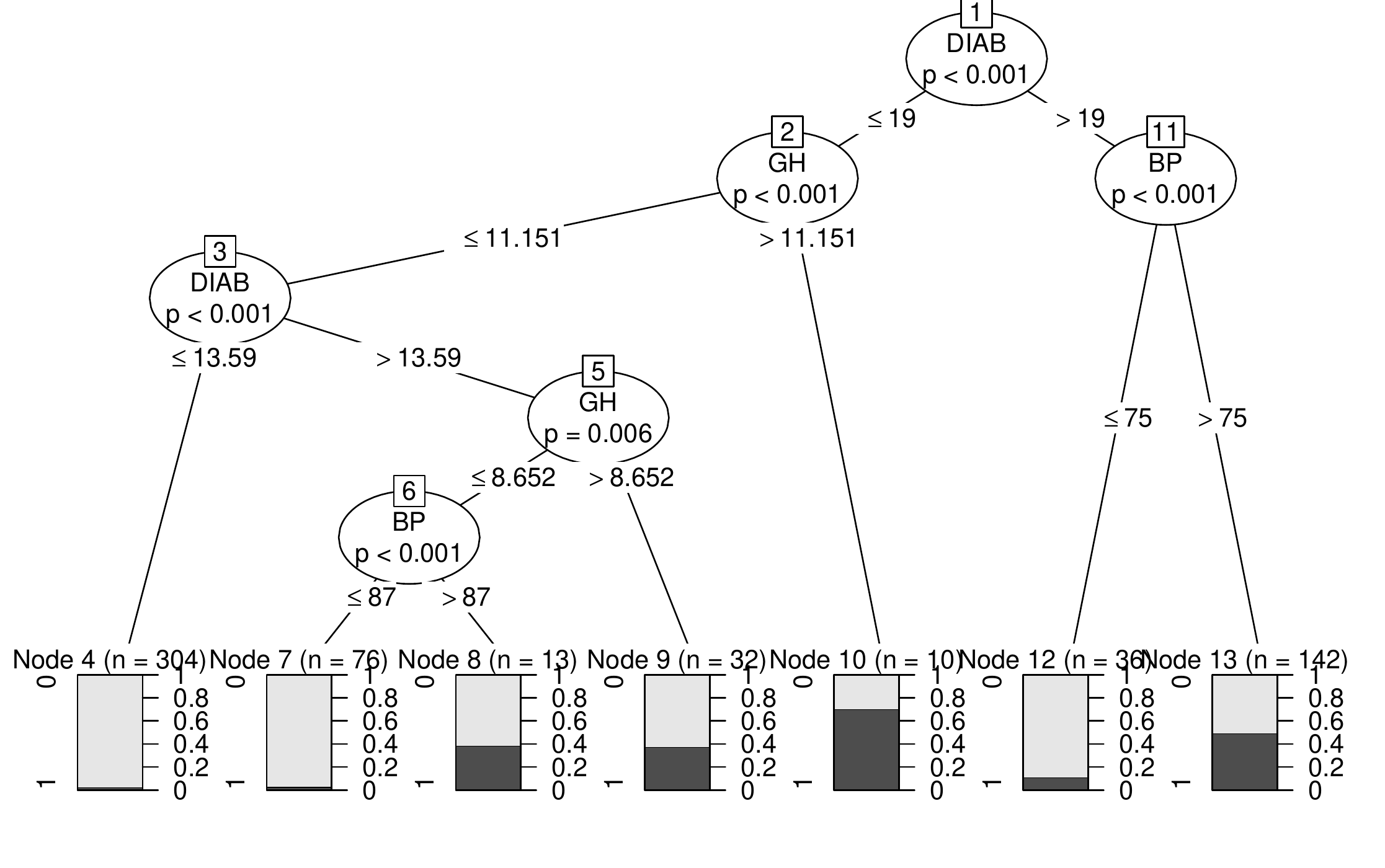}
\caption{Conditional  trees for retinopathy data, upper panel: split between $\{1\}$ and $\{2,3\}$. lower panel: split between $\{1,2\}$ and $\{3\}$.}
\label{fig:retino1}
\end{figure}

%\subsubsection*{Trees for Split Variables}
Split variables are binary and therefore binary trees can be fitted.  Let the tree for $Y_r$   be given by 
\begin{equation}\label{eq:cum}
\log \frac{ P(Y_r=1|\xb )}{P(Y_r=0|\xb)}= \tr_r(\xb),  \quad r=2,\dots,k,
\end{equation}
where $\tr_r(\xb)$ denotes the partitioning of the predictor space, that is, the tree. Then one obtains for the probabilities
\[
P(Y_r=r|\xb )=P(Y \ge r|\xb ) = \frac {\exp(\tr_r(\xb))}{1+\exp(\tr_r(\xb))}.
\]
The corresponding trees are called \textit{split-based trees}. Split variables are a formal tool to group categories but have substantial meaning in many applications. 
For example, in the retinopathy data set \citep{BenGro:98}, which will also be considered later, the response categories are 1: no retinopathy, 2: nonproliferative retinopathy, 3: advanced retinopathy or blind. Thus  the split between categories $\{1\}$ and $\{2,3\}$ distinguishes between healthy and not healthy, whereas the split between $\{1,2\}$ and $\{3\}$ distinguishes between serious illness and otherwise.
It is crucial that explanatory variables may play different roles for different splits. In the retinopathy data set, 
with explanatory variables smoking (SM = 1: smoker, SM = 0: non-smoker), diabetes duration (DIAB) measured in years, glycosylated hemoglobin
(GH),  measured in percent, and diastolic blood pressure (BP) measured in mmHg, one obtains for the two splits the trees shown in Figure \ref{fig:retino1}
(fitted by using \textit{ctree}, \citet{Hotetal:2006}). It is seen that trees are quite different, which means that explanatory variables play differing roles when used to distinguish  between healthy and not healthy and between serious illness and less serious illness. 

%\subsection{Local Modelling:  Adjacent Categories Trees and Random Forests} 
\subsection{Trees for Conditional Splits}

Instead of the unconditional split variables considered previously let us consider the conditional binary variables
\begin{equation}\label{eq:cond3}
\tilde Y_{r}=\left\{
\begin{array}{ll}
1&Y_{ }  \ge r \text{ given } Y_{ }  \in  \{r-1,r\}\\
0&Y_{ } < r \text{ given } Y_{ }  \in  \{r-1,r\},
\end{array} 
\right.
\end{equation}
$r=2,\dots,k$. The variables are conditional versions of split variables. More concrete, $\tilde Y_{r}$ represents $Y_r|Y_{r-1}=1,Y_{r+1}=0$. 
The main difference between $\tilde Y_{r}$ and $Y_{r}$ is that the former is a conditional variable. This is important since fitting a tree to $\tilde Y_{r}$
means one includes only observations with $Y_{ }  \in  \{r-1,r\}$. The corresponding tree can be seen as a nonparametric version of the adjacent categories model and is called an \textit{adjacent categories tree}.  The corresponding trees are local, they reflect the impact of explanatory variables on the distinction between 
adjacent categories.

\begin{figure}[h!]
\centering
\includegraphics[width=0.45\textwidth]{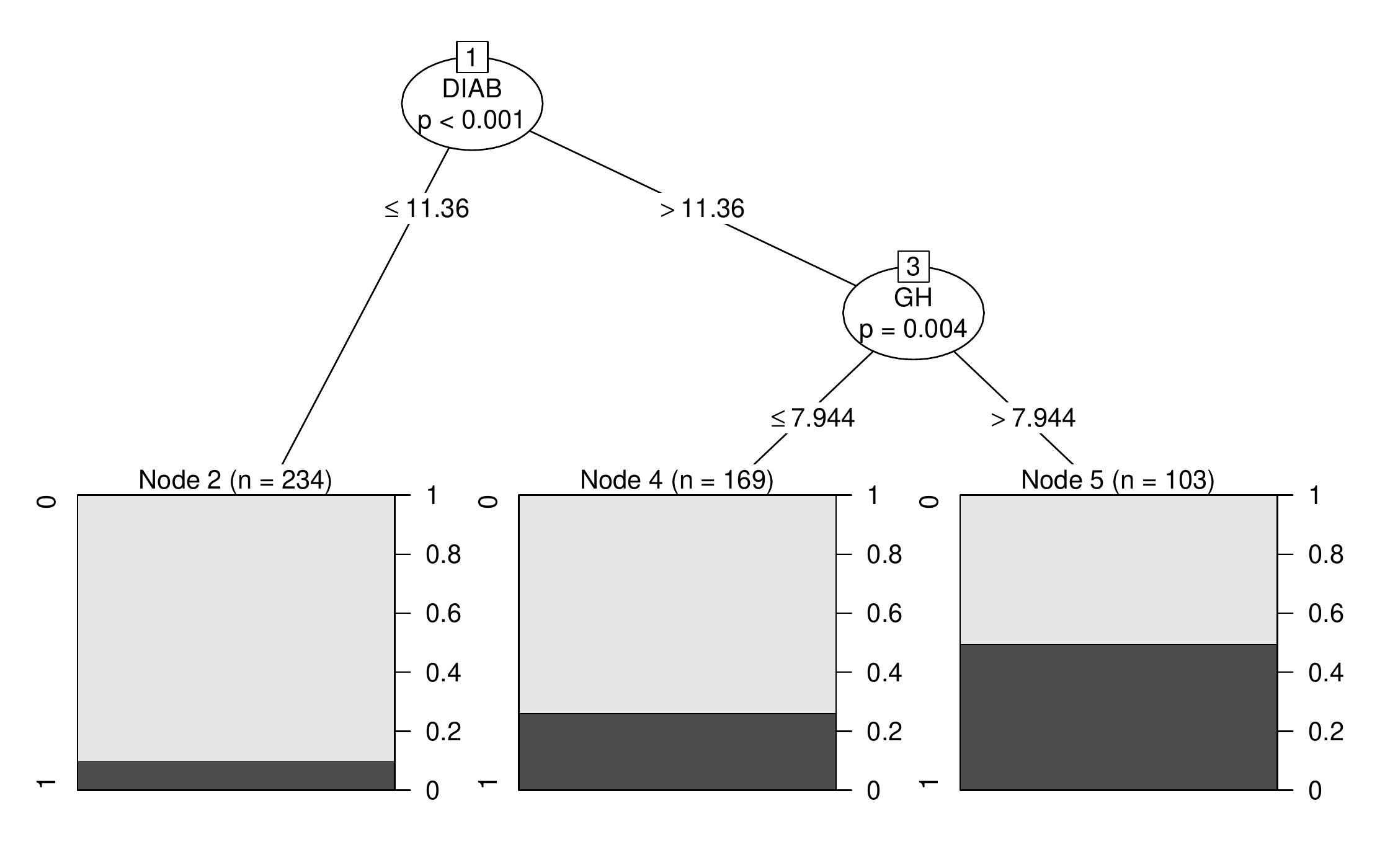}
\includegraphics[width=0.45\textwidth]{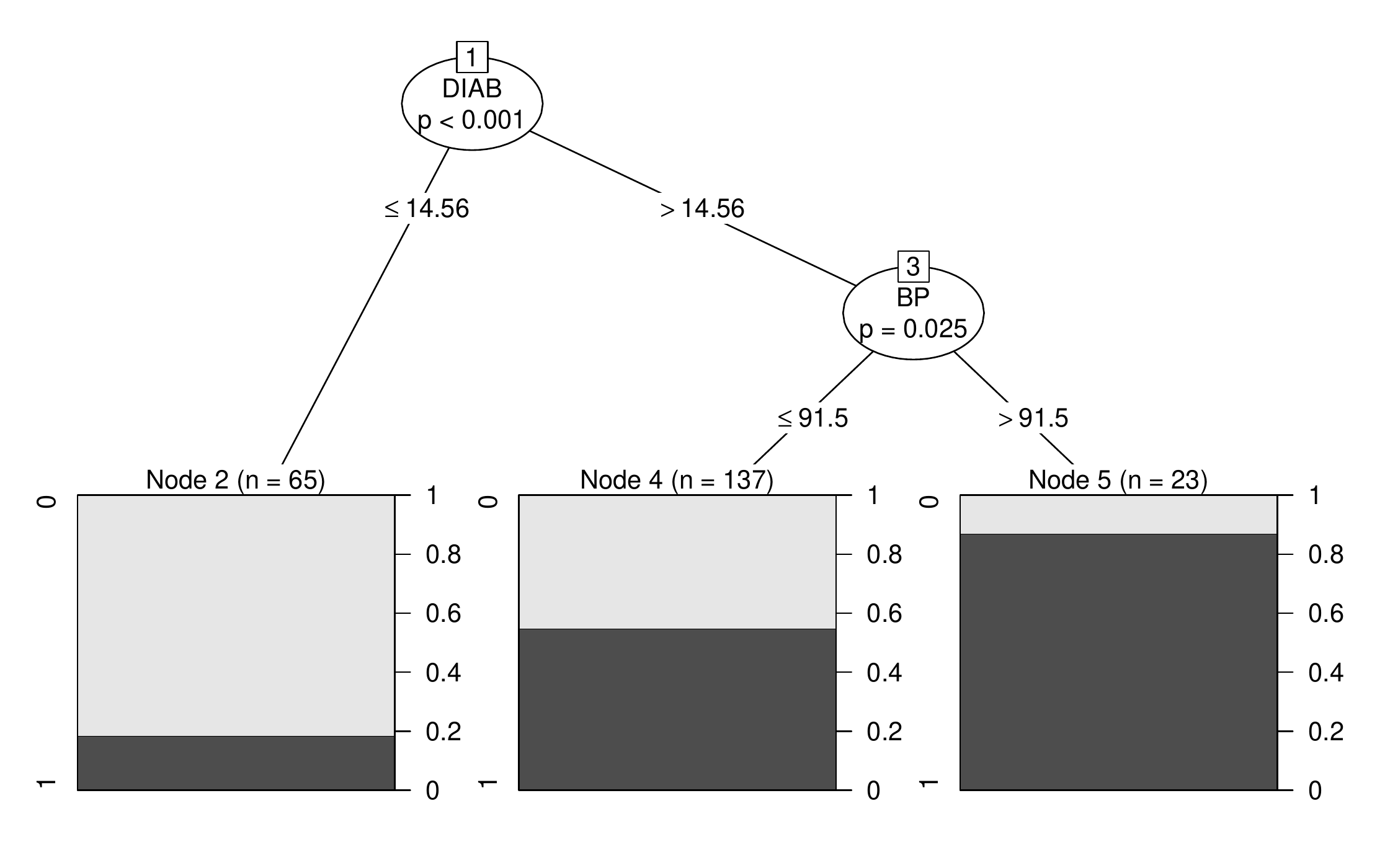}
\caption{Conditional  trees for retinopathy data, left: split given  $Y \in \{1,2\}$, right: split given  $Y \in \{2,3\}$.}
\label{fig:retinoadj}
\end{figure}

Adjacent categories trees have a different interpretation than trees for split variables. For illustration,
Figure \ref{fig:retinoadj} shows the fitted trees for the retinopathy data. It is seen that diabetes duration (DIAB) has an impact in both trees. In the split between categories 1 and 2 the only other variable that is significant is glycosylated hemoglobin while in the split between categories 2 and 3 it is blood pressure. Trees are smaller than  split-based trees since due to conditioning the number of observations is smaller. From a substantial point of view it might be most interesting to combine trees from the different splitting concepts. The first tree in Figure  \ref{fig:retino1} distinguishes between $\{1\}$ and $\{2,3\}$,
that is between healthy and non healthy. The second tree in Figure \ref{fig:retinoadj} shows which variables   are significant when distinguishing between categories 2 and 3 given   the response is in categories $\{2,3\}$, that is, which variables are influential given the patient suffers from retinopathy.

\subsection{From Trees to Random Forests}

Single trees can be informative for researchers that want to investigate which variables have an impact on specific dichotomizations.
If one has prediction in mind a better choice are random trees, which are much more stable and efficient than single trees \citep{Breiman:96, Breiman:2001a,buhlmann2002analyzing}.  Then it is necessary to combine the results of single trees in a proper way. 

Let us first consider split-based trees. They face the problem familiar from cumulative models with category-specific effects that specific constraints have to be fulfilled. More specifically, for all values of $\xb$ the constraint $P(Y \ge 2 |\xb)\ge \dots \ge P(Y \ge k-1 |\xb)$ has to hold, which is equivalent to $P(Y_2 =1|\xb)\ge \dots \ge P(Y_{k-1} =1 |\xb)$.  However, for separately fitted trees the corresponding condition $\tr_2(\xb)\ge \dots \ge \tr_{k-1}(\xb)$ does no necessarily hold. The same problem occurs in partial proportional odds model, for which   $\beta_{02}+\xb^T\betab_2\ge \dots \ge \beta_{0k}+\xb^T\betab_k$ has to hold.

Let $\hat\pi(\xb)^{(r)}=\hat P(Y \ge r|\xb)$ denote the estimated cumulative probabilities resulting from the tree for the split variable $Y_r$. Then probabilities are obtained by $\hat P(Y = r|\xb)= \hat\pi(\xb)^{(r)} -\hat\pi(\xb)^{(r+1)}$ if $\hat\pi(\xb)^{(r)} \ge \hat\pi(\xb)^{(r+1)}$ for all $r$. If the latter condition does not hold cumulative probabilities $\hat\pi(\xb)^{(r)},\dots, \hat\pi(\xb)^{(k)}$  are fitted to be decreasing  by using monotone regression tools.  Alternative approaches to obtain compatible estimators have been considered in the machine learning community, for example, by \citet{chu2007support}.

An advantage of adjacent categories trees is that no monotonization tools are needed since estimated probabilities are always compatible.
Let the adjacent categories trees be given by 
\begin{equation}\label{eq:adj}
\log \frac{ P(\tilde Y_r=1|\xb )}{P(\tilde Y_r=0|\xb)}= \tilde\tr_r(\xb),  \quad r=2,\dots,k,
\end{equation}
where $\tilde\tr_r(\xb)$ denotes the partitioning of the predictor space. It is not hard to derive that  the probability of an response in category $r$ given the representation (\ref{eq:adj}) holds has the form
\begin{equation}\label{eq:comb}
P(Y_r=r|\xb )= \frac{\exp(\sum_{s=2}^r \tilde\tr_s(\xb))}{\sum_{s=1}^k\exp(\sum_{l=2}^s \tilde\tr_s(\xb))},
\end{equation}
where $\sum_{l=2}^1 \tilde\tr_s(\xb)) =0$. The representation (\ref{eq:comb}) holds for any values of $\tilde\tr_2(\xb),\dots, \tilde\tr_{k}(\xb)$, no specific restriction has to be fulfilled.

Random forests are obtained by combining not only the trees for split variables but averaging over a multitude of  trees generated by randomization.
The approach proposed here  exploits the role of the split variables as building blocks for ordinal responses, and can be seen as a \textit{split variables based} approach, which is unconditional in split-based trees and conditional in adjacent categories trees. 

\subsection{Ensemble Learners Including Parametric Models} 

Before investigating the proposed random forests in detail let us point out a problem with ordinal trees that is often ignored. 
Most presentations of ordinal trees focus on the development of novel trees but do not compare the performance of random forests to the performance of 
simple parametric models as the proportional odds model. That leaves the impression that random forests are the most efficient tools. As will be demonstrated in the following sections parametric models should not be ignored, in many applications they can perform as well as random forests or even better.
The use of parametric ordinal models for the prediction of ordinal responses has some tradition, see, for example, \citet{Rud-etal:95}, \citet{CamDon:89}, \citet{Cam-etal:91}, \citet{AndPhi:81}.

Trees  themselves are ensemble methods that combine various trees to obtain a good approximation of the underlying response probabilities. To exploit the potential strength of parametric models we propose an ensemble that includes these models. When estimating response probabilities we will use the ensemble
\[
\hat P(Y =r|\xb) = \sum_{j=1}^M  w_j \hat P_j(Y =r|\xb),
\]
where $\hat P_j(Y =r|\xb)$  are estimated probabilities for the $j$-th learner. Learners can be random forests but also parametric models. The weights $w_j$
are chosen according to the prediction performance of the $j$-th learner. The ensemble efficiently uses different types of learners. By combining them it yields more stable predictions than  single learners and automatically gives more weight to the best learner in the ensemble. 

Typically, in classification predictions of single trees from an ensemble are combined by voting. Each subject with given values of the predictor is dropped through every tree such that each single tree returns a predicted class. The prediction of the ensemble is the class  most trees voted for. One obtains a majority vote , which has also been called a committee method.  It should be noted that the ensembles proposed here combine probabilities. They are not ensembles that use majority votes  to combine class predictions obtained for each single learner. By computing the predicted class probabilities on can use more general accuracy measures that also take into account  the precision of the prediction. Moreover, with ordinal responses it is sensible not to use the mode of the class probabilities but the median computed over the predicted probabilities as predicted class. The use of estimated probabilities is of crucial importance. We also considered majority votes that combine the votes on splits but the results were distinctly inferior to using probabilities.

\section{Ordinal Random Forests and Prediction}\label{sec:pred} 

\subsection{Measuring Accuracy of Prediction } 
One way to investigate the power of a model is to investigate its ability to predict future observations. In discriminant analysis one often uses class prediction as a measure of performance. Class prediction in the considered framework comes in two forms. As predicted class one may use the mode of the response,  $\hat Y =\mod(\xb)$, which is in accordance with the Bayes prediction rule, or  the median $\hat Y=\med(\xb)$, which makes use of the ordering of categories. Then for a new observation $(Y_0,\xb_0)$ one typically considers the 0-1 loss function 
\[
L_{01}(Y_0,\hat Y_0)= I(Y_0 \ne \hat Y_0),
\]
where $I(.)$ is the indicator function. One obtains 1 if the prediction is wrong, and 0 if the prediction is correct. The average over new observations yields the 0-1 error rate.

Rather than giving  just one value as a predictor for the class it is more appropriate to consider the whole vector $\hat\pib^T(\xb)=(\hat\pi_1(\xb), \dots,\hat\pi_k(\xb))$, where $\hat\pi_r(\xb)=P(Y_r=r|\xb)$ is the probability one obtains after fitting a tree. The vector $\hat\pib(\xb)$ represents the predictive distribution. 
As \citet{Gneitingetal:2007} postulated a desirable predictive distribution should be as sharp as possible and
well calibrated. Sharpness refers to the concentration of the distribution and calibration to the agreement between distribution
and observation.

Since the response is measured on an ordinal scale  an appropriate loss function  derived from the
continuous ranked probability score (\citet{Gneitingetal:2007}) is
\[
L_{RPS}(Y_0,\hat{\pib})=\sum_{r=1}^k(\hat\pi(r,\xb_0)- I(Y_0 \le r))^2,
\]
where $(Y_0,\xb_0)$ is a new observation and $\hat\pi(r,\xb_0)=\hat\pi_1(\xb_0)+\dots+\hat\pi_r(\xb_0)$ is the cumulative probability. It is a sum over
quadratic (or Brier) scores for binary data and takes the closeness between the whole distribution and the observed value into account.
For alternative measures see also \citet{Gneitingetal:2007}.

\subsection{Data Sets }

\subsubsection*{Heart Data }
This data set includes 294 patients undergoing angiography at the Hungarian Institute of Cardiology in Budapest between 1983 and 1987, and is included in the R package \textit{ordinalForest} \citep{hornung2019ordinal}. It contains ten covariates and one ordinal target variable. Explanatory variables are 
age (age in years), sex (1 = male; 0 = female), chest pain (1 = typical angina; 2 = atypical angina; 3 = non-anginal pain; 4 = asymptomatic),
trestbps (blood pressure in mm Hg on admission to the hospital),  chol (serum cholestoral in mg/dl), fbs (fasting blood sugar $>$ 120 mg/dl, 1 = true; 0 = false)
restecg (resting electrocardiographic results, 1 = having ST-T wave abnormality, 0 = normal), thalach (maximum heart rate achieved),
exang (exercise induced angina, 1 = yes; 0 = no),
oldpeak (ST depression induced by exercise relative to rest). The response is
Cat (severity of coronary artery disease determined using angiograms, 1 = no disease; 2 = degree 1; 3 = degree 2; 4 = degree 3; 5 = degree 4).

\subsubsection*{Birth Weight Data }
The \textit{lobwt} data set contained in the R package \textit{rpartOrdinal}  has been used  in  several random forest papers. As categorical response we  use the birth weight by binning the variable bwt according to the cutoffs: 2500,3000, and 3500 (see also  \citep{galimberti2012classification}). Explanatory variables are 
age (age of mother in years), lwt  (weight of mother at last menstrual period in Pounds),
smoke (Smoking status during pregnancy, 1: No, 2: Yes), 
ht (history of hypertension, 1: No, 2: Yes),
ftv (number of physician visits during the first trimester,1: None, 2: One, 3: Two, etc)

\subsubsection*{GLES Data }
The GLES  data stem from the German Longitudinal Election Study (GLES), which is a long-term study of the German electoral process \citep{GLES}.  The data consist of  $2036$  observations and  originate from the pre-election survey for the German federal  election  in  2017 and are concerned with political fears. In particular the participants were   asked: ``How afraid are you due to the use of nuclear energy?
The answers were measured on Likert scales from 1 (not afraid at all) to 7 (very afraid). The  explanatory variables in the model are
\textit{Abitur} (high school leaving certificate,  1: Abitur/A levels; 0: else),
\textit{Age} (age of the participant),
\textit{EastWest} (1: East Germany/former GDR; 0: West Germany/former FRG),
\textit{Gender} (1: female; 0: male),
\textit{Unemployment} (1: currently unemployed; 0: else).

\subsubsection*{Safety Data }
The package CUB \citep{iannario2018cub} contains the data set relgoods, which provides results of a survey aimed at measuring the subjective extent of feeling safe in the streets. The data were collected  in the metropolitan area of Naples, Italy. Every participant was asked to assess on a 10 point ordinal scale his/her personal score for  feeling  safe with large categories referring to feeling  safe.
There are $n=2225$ observations and five variables,  \textit{Age}, \textit{Gender} (0: male, 1: female),   the educational degree (\textit{EduDegree}; 1: compulsory school, 2: high school diploma, 3: Graduated-Bachelor degree, 4: Graduated-Master degree, 5: Post graduated),
\textit{WalkAlone} (1 = usually walking alone, 0 = usually walking in company), 
\textit{Residence} (1: City of Naples, 2: District of Naples, 3: Others Campania, 4: Others Italia).

\subsection{Ensembles at Work }
In the following the accuracy of prediction in the data sets described above  is investigated. The data sets were split repeatedly into a learning set with sample size $n_L$ and a validation set built from the rest of the data.
 (number of splits: 30). The learning set was used to fit the method under investigation, the accuracy of prediction is then computed in the validation set.
We use the ranked probability score, since it indicates the accuracy of prediction better than simple class predictions. In addition, we give, for simplicity, the distance between the predicted class and the true class. The latter measure is an indicator how far the prediction is from the true class.

\begin{figure}[h!]
\centering
\includegraphics[width=0.45\textwidth]{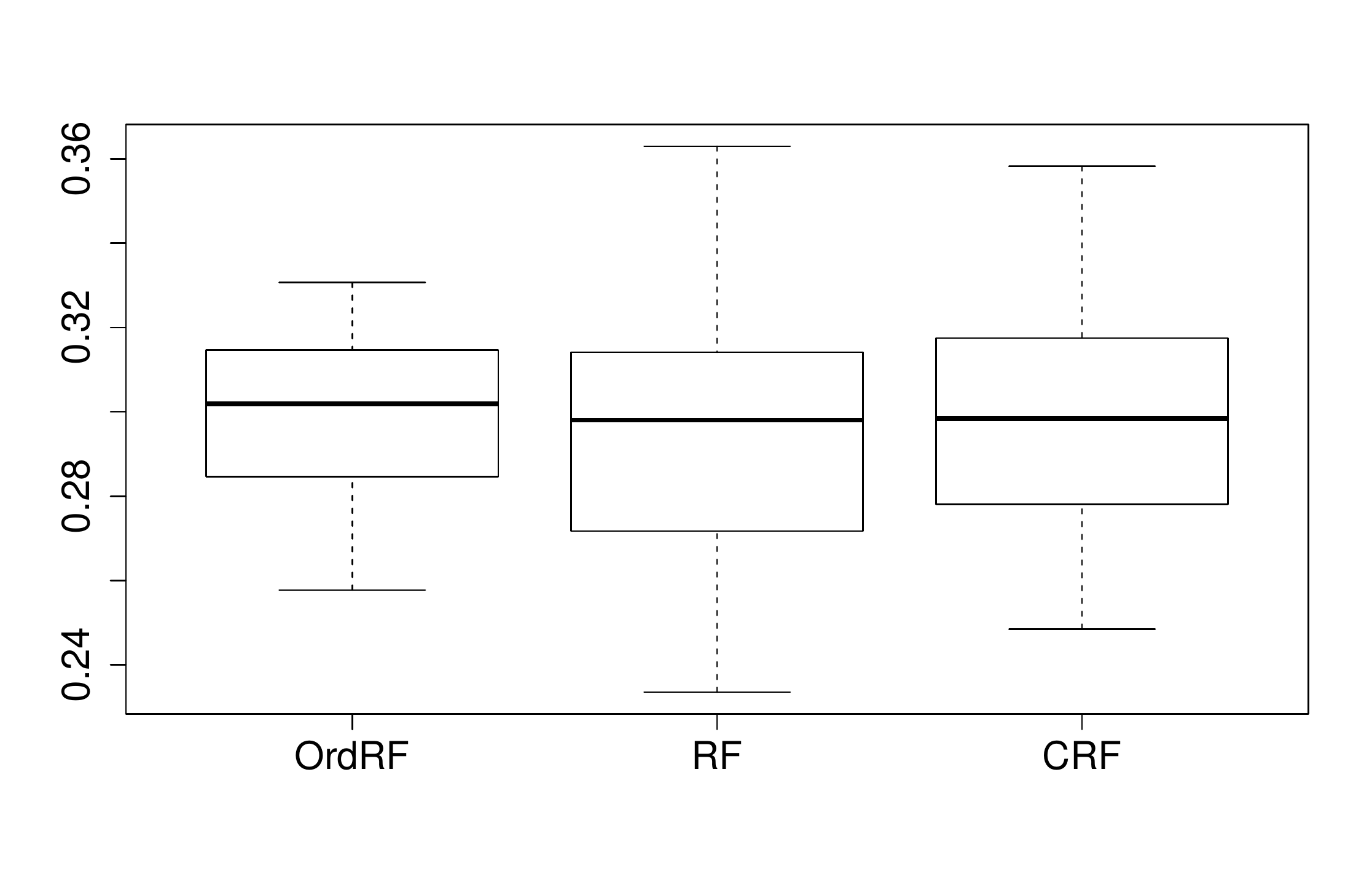}
\includegraphics[width=0.45\textwidth]{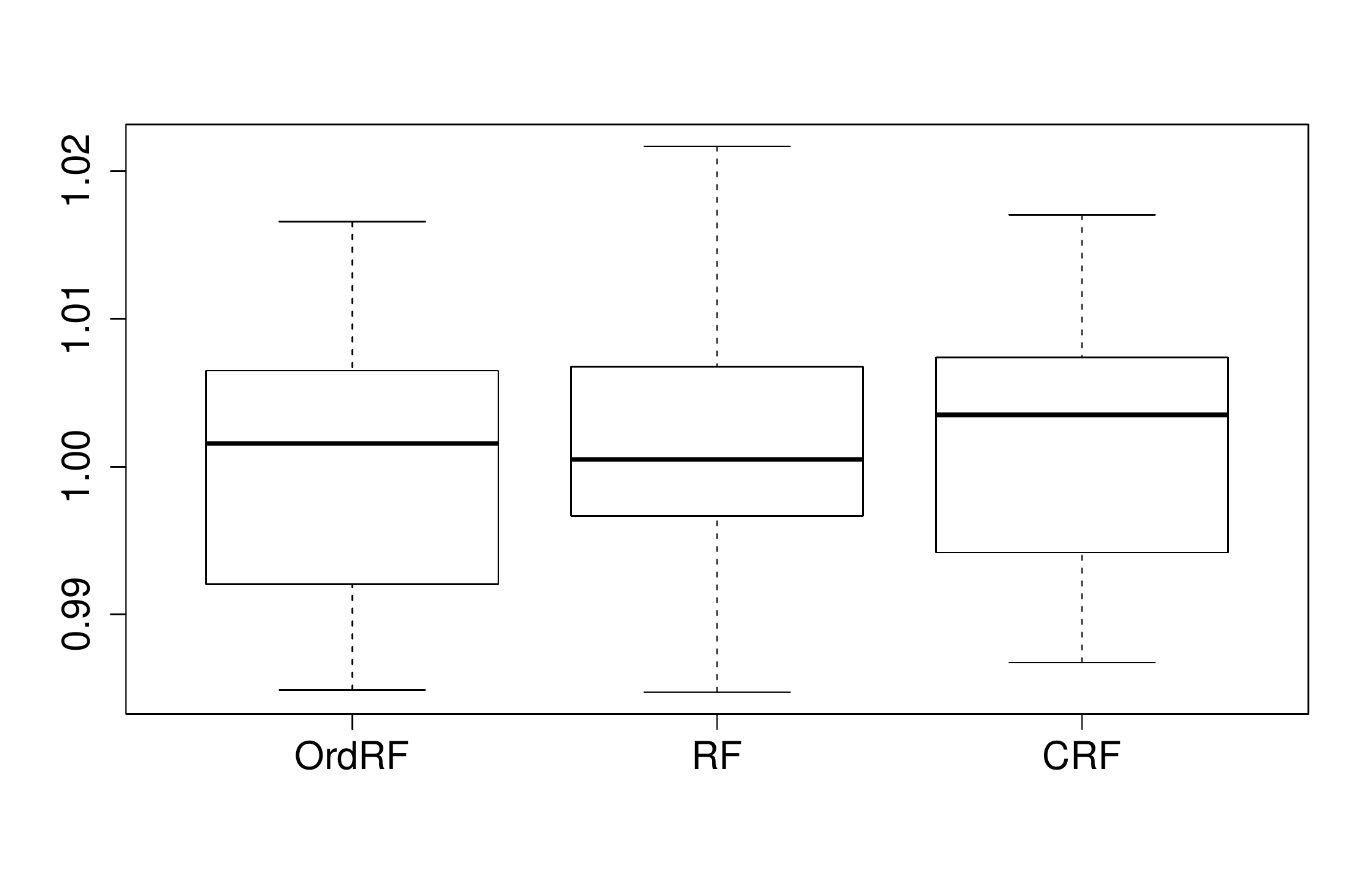}
\caption{Ranked probability scores for housing data (left) and GLES data (right) when fitting the adjacent categories random forest with ordinalForest (OrdRF), randomforest (RF), and cforest (CRF).}
\label{Houscomp}
\end{figure}

The fitting of split-based and adjacent categories random forests can be based on different random forest methods for binary responses. In particular one can use
\textit{ordinalForest} \citep{hornung2019ordinal}, \textit{randomforest} \citep{liaw2015package}, or conditional trees as provided by \textit{cforest} \citep{hothorn2015partykit}. Figure  \ref{Houscomp}  shows the averaged ranked probability scores for fitted adjacent categories random forests when using ordinalForest (OrdRF), randomforest (RF), and cforest (CRF) for the housing data and the GLES data. The differences in performance are negligible. Therefore, in the following we use only one method to generate split-based and adjacent categories random forests, namely \textit{randomforest}, which is computationally quite efficient. 
%Since ordinalForest is one of the competitors, which is considered because it does not use fixed assigned scores, we also use ordinalForest to generate the binary trees that are used to obtain the random forests. 

The methods to be considered in the following are:
\begin{itemize}                 
\item   \textit{Pom}: fitting of a proportional odds model,
\item  \textit{Adj}: fitting of an adjacent categories logit model,
\item   \textit{RFord}: fitting of an ordinal forest with \textit{ordinalForest}, 
\item   \textit{RFSplit}: Split-based ordinal random forest using \textit{randomForest} to fit the binary random forests,
\item   \textit{RFadj}: fitting of an adjacent categories random forest using \textit{randomForest} to fit the binary random forests,
\item   \textit{Ens3}: weighted ensemble including the proportional odds model, ordinalForest fit and adjacent categories random forest,
\item   \textit{Ens5}: weighted ensemble including the proportional odds model, the adjacent categories model, ordinalForest fit,and adjacent categories and split-based random forest.
\end{itemize}

The last two methods are ensemble methods that include parametric models. \textit{Ens3} is built from one parametric model, an ordinal random forest, and the adjacent categories random forest, whereas  \textit{Ens5} contains in addition the adjacent categories model and the split-based random forest.
The ensemble built from three methods serves to demonstrate that it is essential to combine ordinal random forests and parametric models. The inclusion of further models will be shown to improve the performance only slightly.

Figure \ref{fig:datasets} shows the ranked probability scores (left column) and the distance between true response and predicted response (right column) for the validation data. It is seen that ordinal random forests outperform parametric models for the first three data sets only. A distinct advantage of ordinal forests is seen in particular in the housing data set. For the heart and birth weight  data sets the performance of ordinal trees is only slightly superior.  For the retinopathy  and the GLES data simple parametric models perform much better than ordinal forests.  The best performance is seen for the ensemble methods that combine parametric and nonparametric methods. They seem to efficiently combine the best of two worlds yielding small errors for all data sets. Thus, if one wants to avoid ending up with an inferior prediction tool one should consider not only trees or parametric models but aim at a combination of these methods.

In Figure \label{fig:datasets} the split-based based and adjacent categories forest utilize the \textit{randomForest} method to fit the contained binary trees. 
Very similar performance is found when using alternative methods to fit binary trees, like \textit{ctree} or \textit{ordinalForest}. These alternative methods yield different  trees. When investigating single trees the choice of the method definitely makes a difference, and specific trees may offer advantages, for example  conditional trees , which use tests in the splitting procedure, are able to control the significance level and avoid selection bias   \citep{strobl2007bias, Hotetal:2006} making them an attractive choice. However, for ensembles of trees as random forests the performance is very similar, at least in the case of split-based based and adjacent categories forests.

\begin{figure}[H]
\centering
\includegraphics[width=0.45\textwidth]{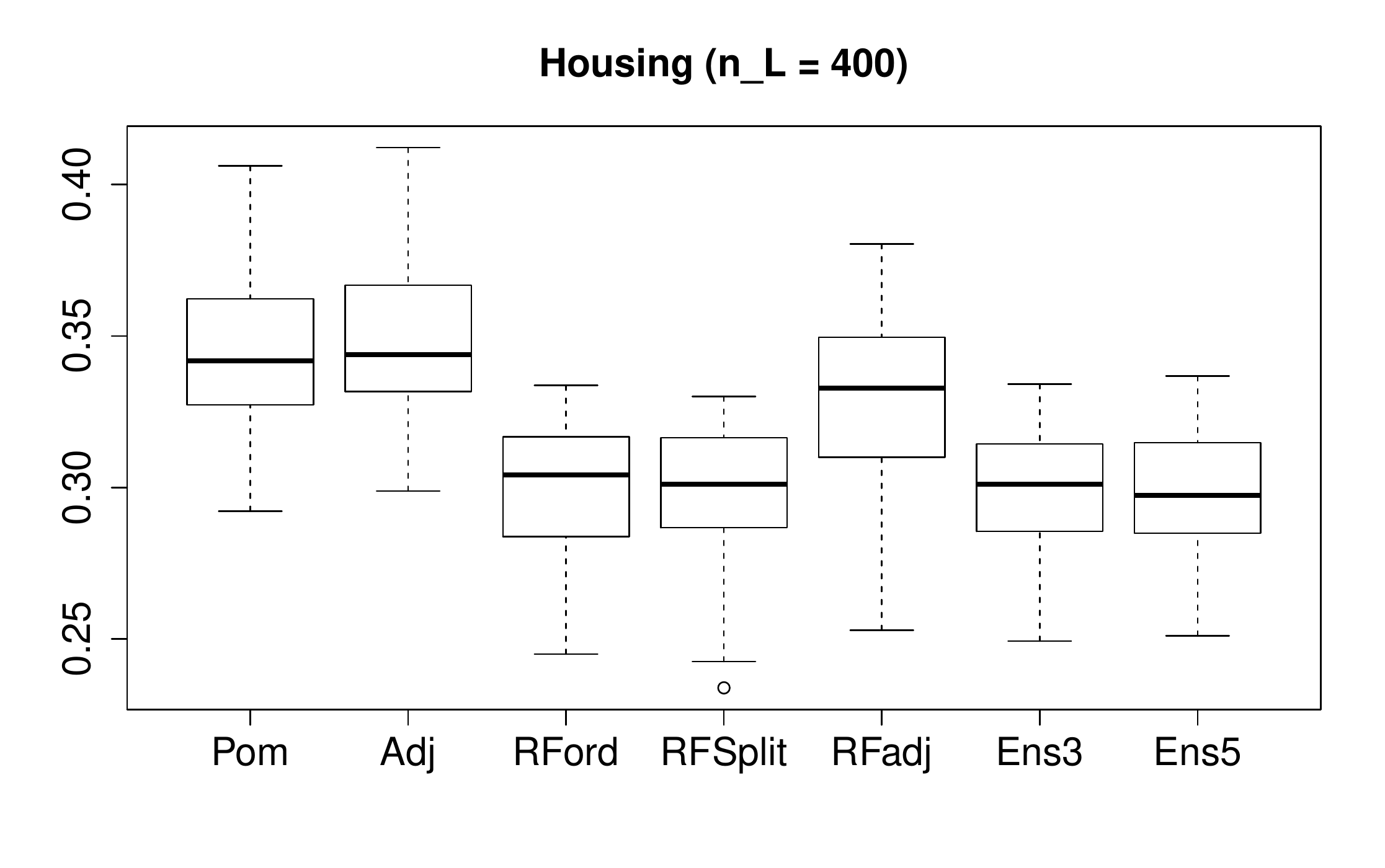}
\includegraphics[width=0.45\textwidth]{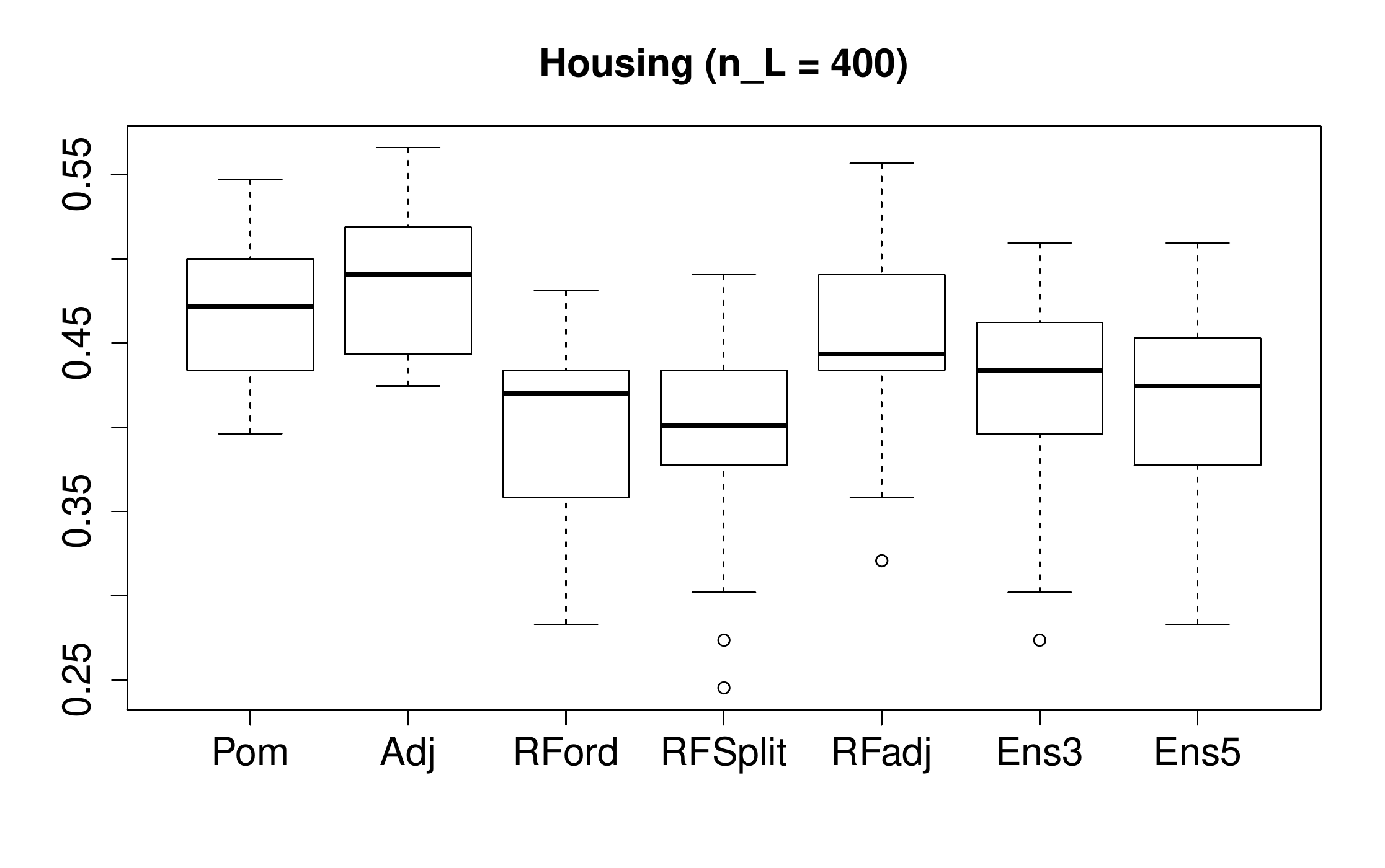}
\includegraphics[width=0.45\textwidth]{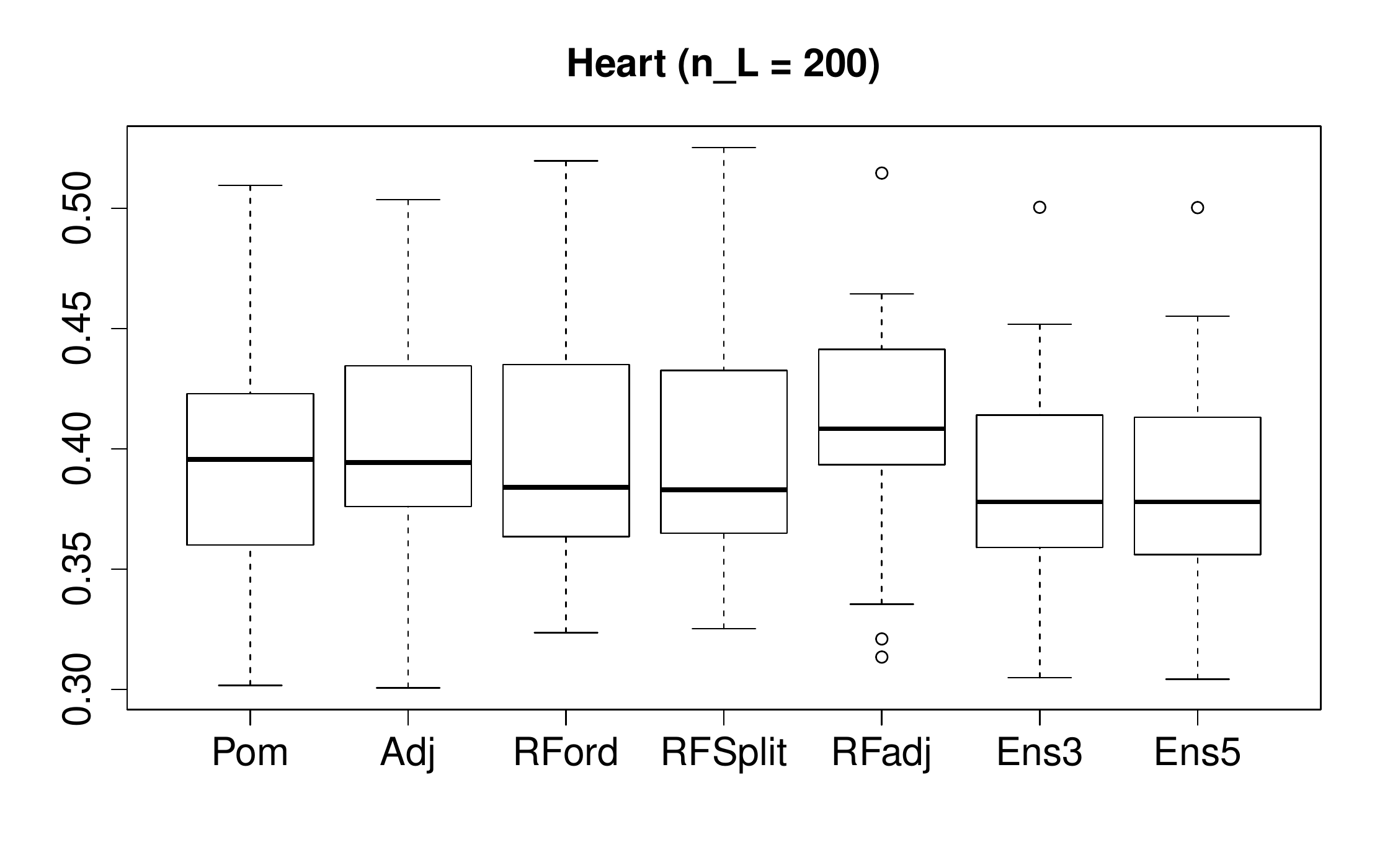}
\includegraphics[width=0.45\textwidth]{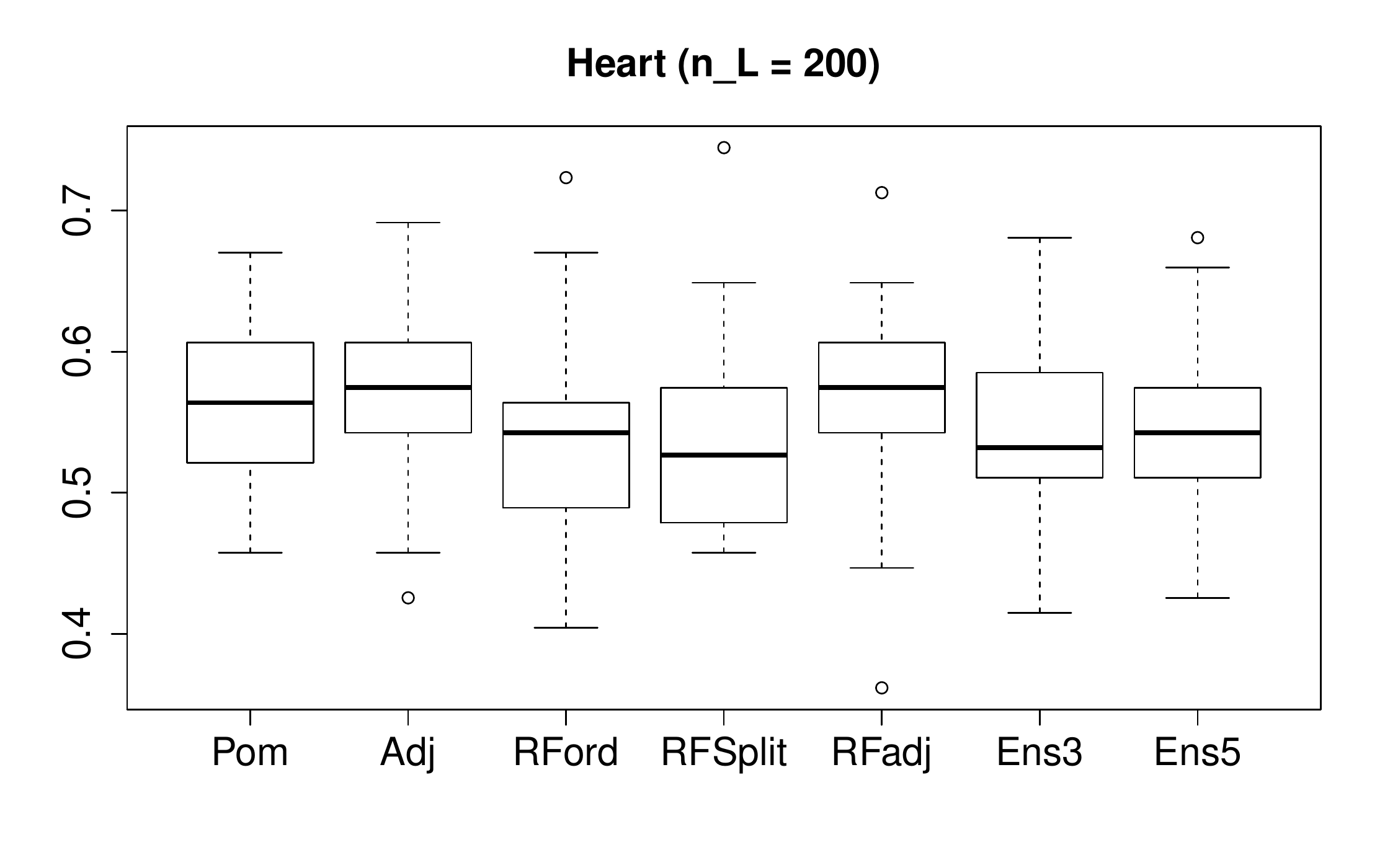}
\includegraphics[width=0.45\textwidth]{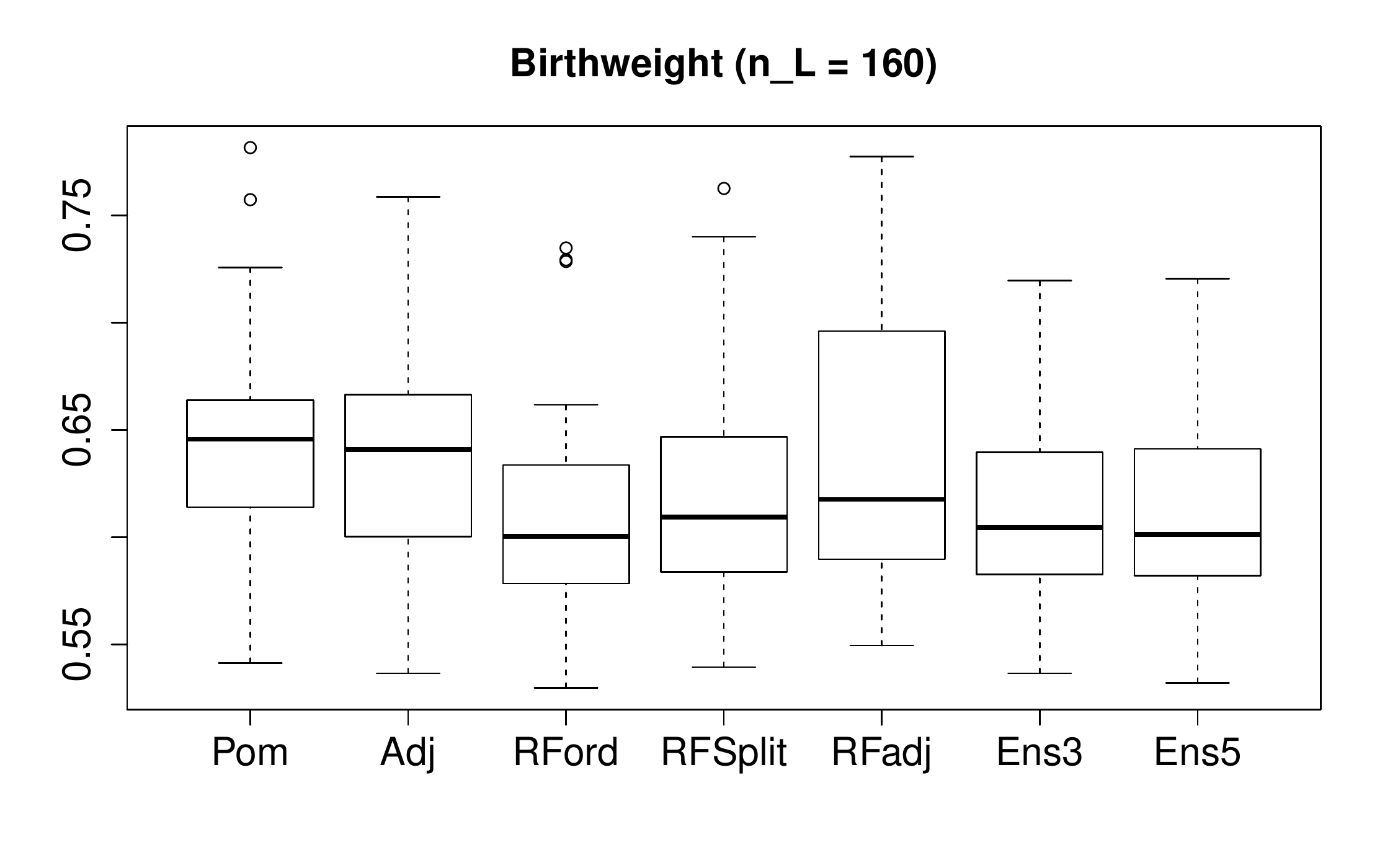}
\includegraphics[width=0.45\textwidth]{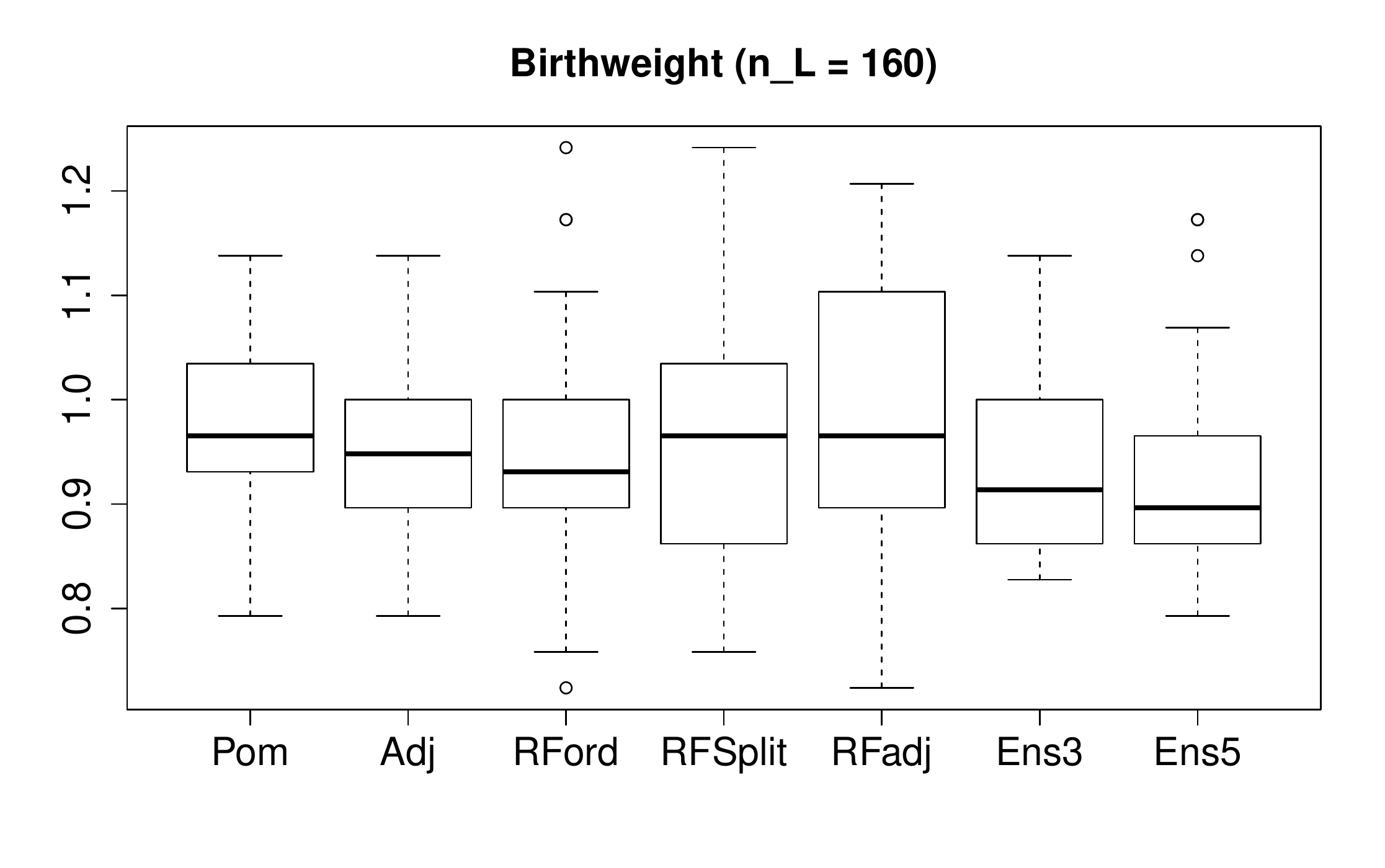}
\includegraphics[width=0.45\textwidth]{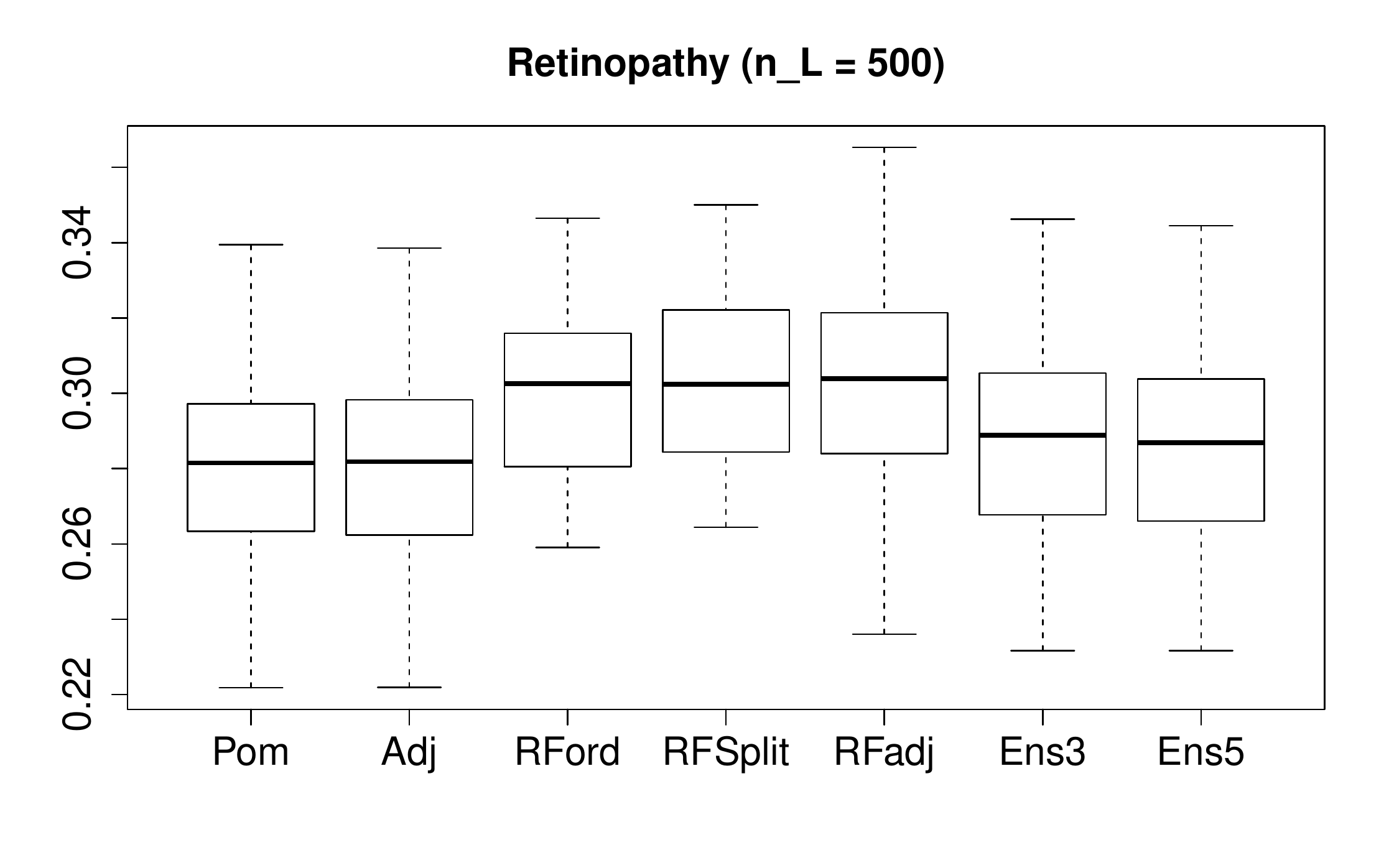}
\includegraphics[width=0.45\textwidth]{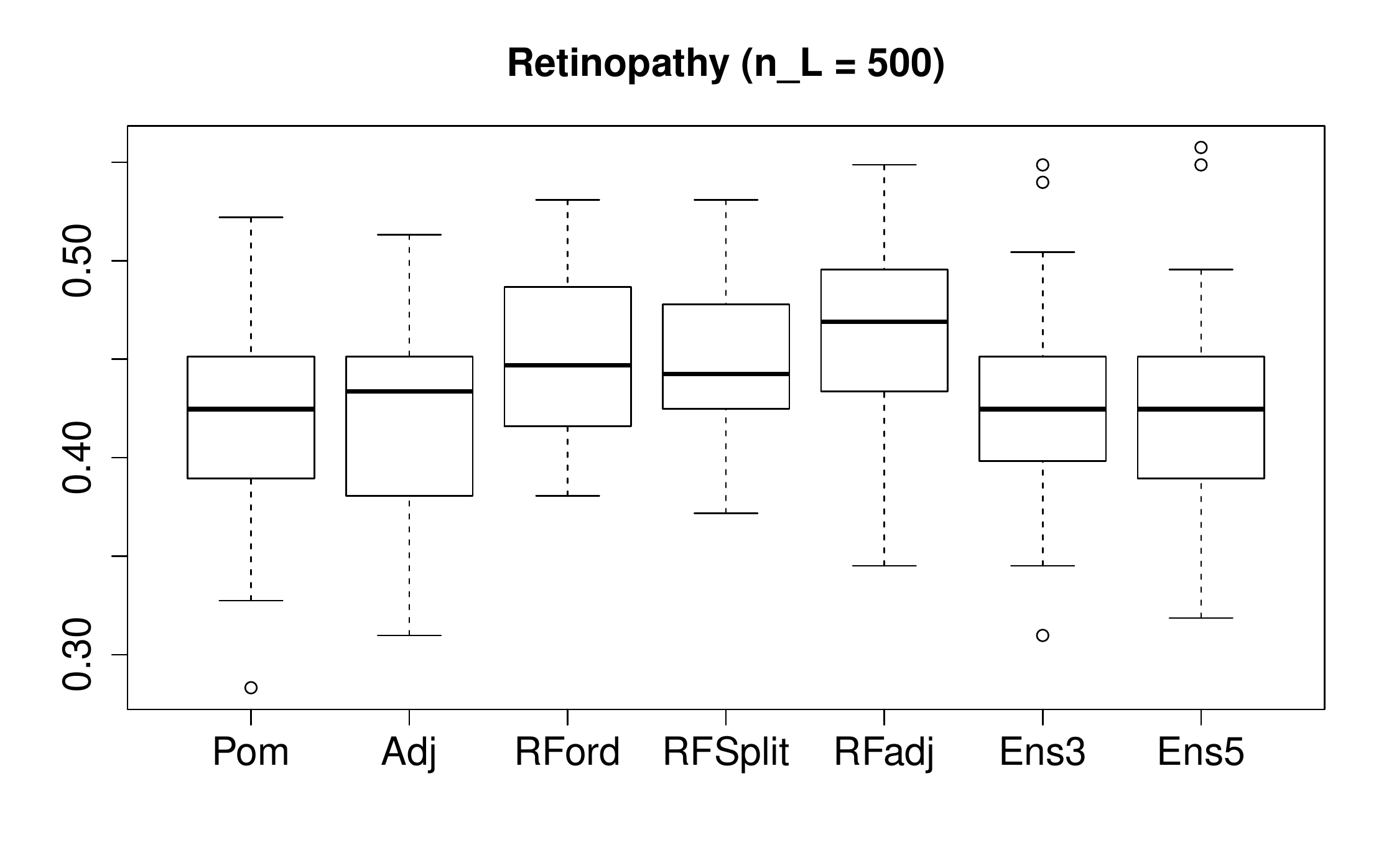}
\includegraphics[width=0.45\textwidth]{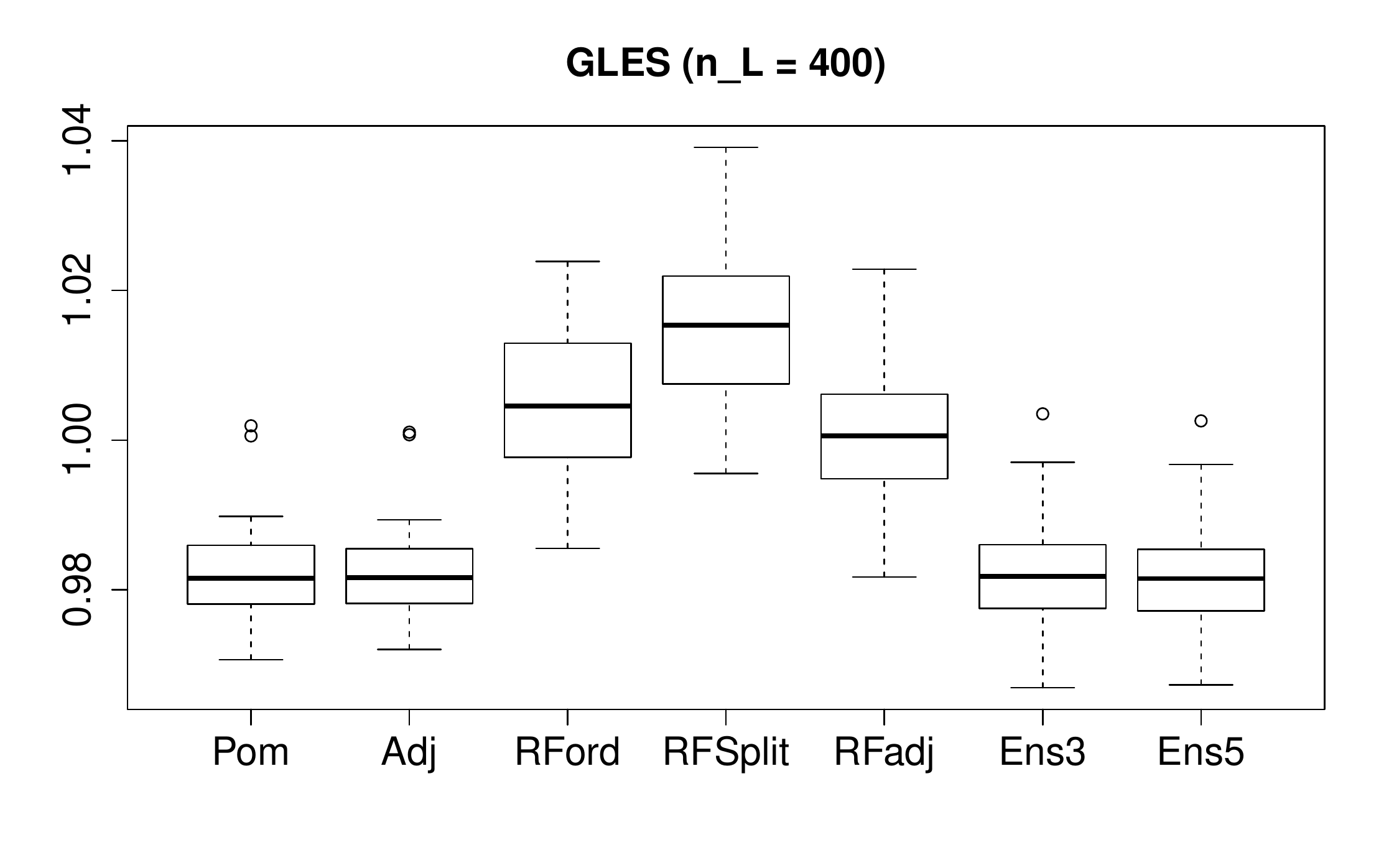}
\includegraphics[width=0.45\textwidth]{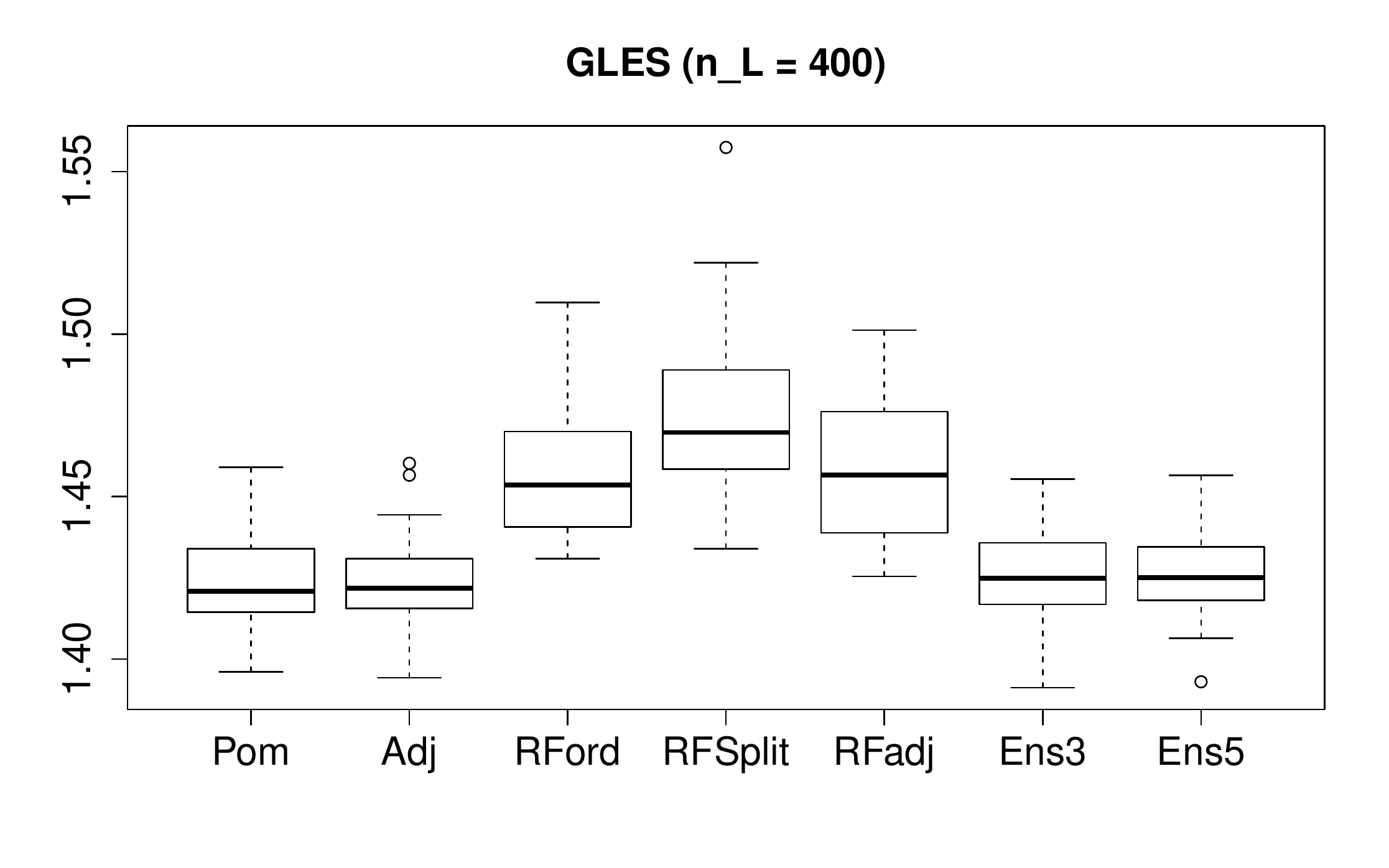}
\caption{Results for data sets; left: ranked probability score, right: distance between true and estimated value. }
\label{fig:datasets}
\end{figure} 
\section{Importance of Variables}\label{sec:imp}

While single trees for split variables are easy to interpret this does not hold for ensembles of trees. Since variables appear in different trees at different positions the impact of variables is hard to infer from plots of hundreds of trees. On the other hand random forests allow for complex effects of predictors, which makes it a flexible prediction tool. 

There is a considerable amount of literature that deals with the development of importance measures for random forests, see, for example, \citet{strobl2007bias,strobl2008conditional,hapfelmeier2014new,gregorutti2017correlation,hothorn2015partykit}.
A naive measure  simply  counts the number of times each variable is selected by the individual trees in the ensemble. 
Better, more elaborate variable importance measures incorporate
a (weighted) mean of the individual trees' improvement
in the splitting criterion produced by each variable.
An example for such a measure  is the "Gini importance" available in the randomForest package.
It describes the improvement in the "Gini gain" splitting criterion. Alternative, and better 
variable importance measures are based on permutations yielding so-called permutation accuracy importance measures \citep{strobl2007bias}.
By randomly permuting single predictor variables $X_j$, the original
association with the response $Y$ is broken. When the permuted
variable $X_j$, together with the remaining un-permuted
predictor variables, is used to predict the response,
the prediction accuracy is supposed to decrease   if the 
variable $X_j$ had an additional impact on explaining the response. The difference
in prediction accuracy before and after permuting $X_j$ yields a permutation accuracy importance measure. 

\begin{figure}[H]
\centering
\includegraphics[width=7cm]{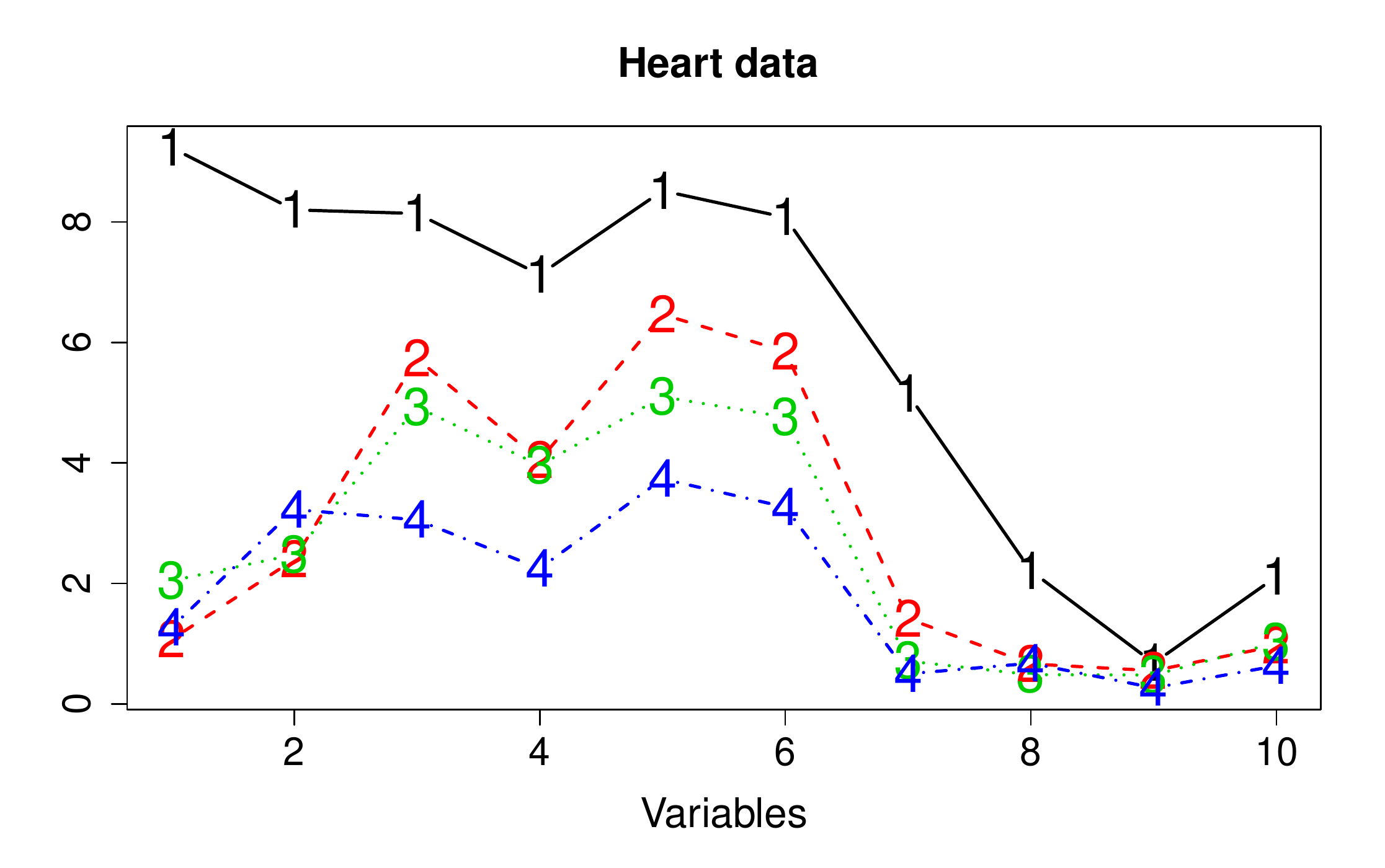}
\includegraphics[width=7cm]{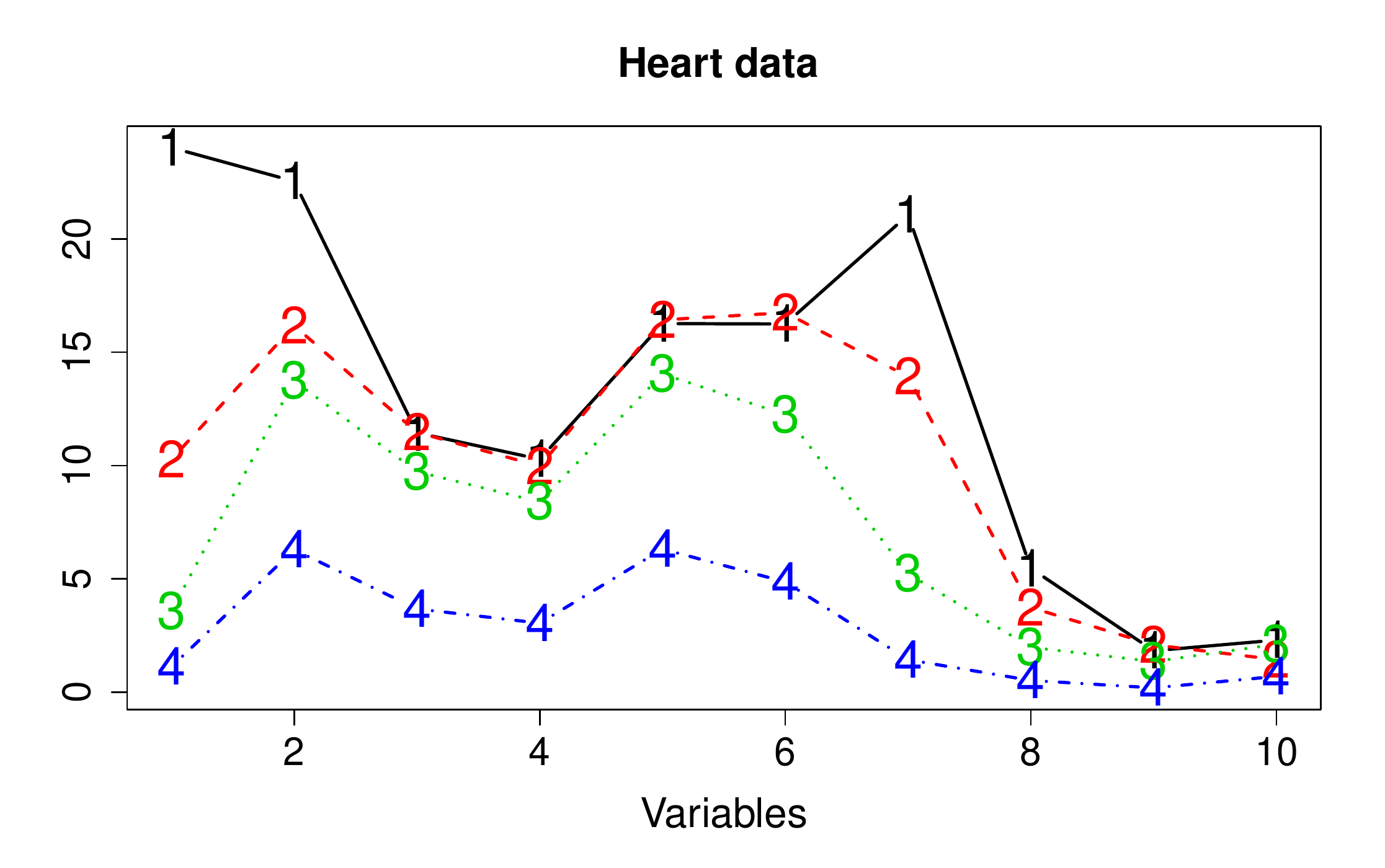}
\includegraphics[width=7cm]{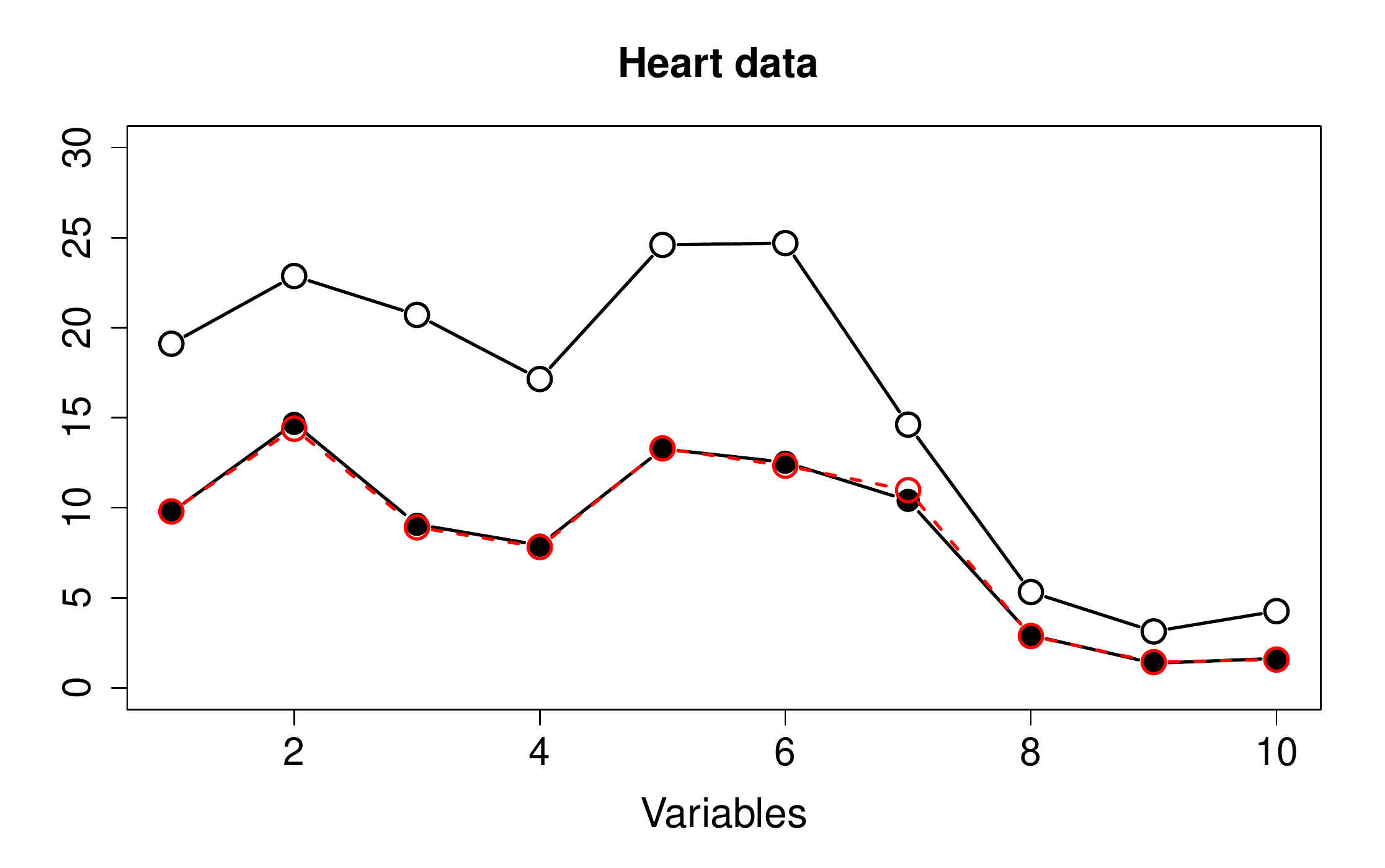}
\caption{Gini importance for heart data; variables 1 to 10: chest pain, oldpeak, age, trestbps, chol, thalach, exang, sex, fbs, restecg (randomForest fit); left upper panel: importance for conditional splits in adjacent categories RF, right upper panel: importance for  splits in split-based RF, lower panel: averaged importance measures for splits in adjacent categories and split-based RF (lower curves) and multi-categorical fit of randomForest (upper curve). }
\label{heartGini}
\end{figure}

In the following we use the heart data to illustrate how importance measures can be obtained for split-based and adjacent categories random forests.
Of course it depends on the algorithm that is used to grow binary trees which importance measure can be computed.   Figure \ref{heartGini} shows the Gini importance when using randomForest to fit the binary random forests. In the upper panels one sees the importance measures obtained for the split variables, that is, for conditional splits in adjacent categories RF on the left, and direct splits for split-based RF on the right.  The numbers $1$ to 4 indicate the  splits. For example, 3 means that the split is between categories $\{1,2,3\}$ and $\{4\}$. 
It is seen that the first six variables show strong importance with the importance being stronger for lower categories splits and weaker for higher category splits. The lower panel  shows the  importance measures averaged across the splits. The lower curves, which are almost identical, show the average for the adjacent categories and split-based random forest. It shows, in addition, the Gini importance for the multi-category random forest obtained from randomForest.
It is seen that the importance measures have the same order for all the fitted random forests. That the values of importance for the multi-category random forest
is higher than for the other two forests is merely a scaling effect.

\begin{figure}[H]
\centering
\includegraphics[width=7cm]{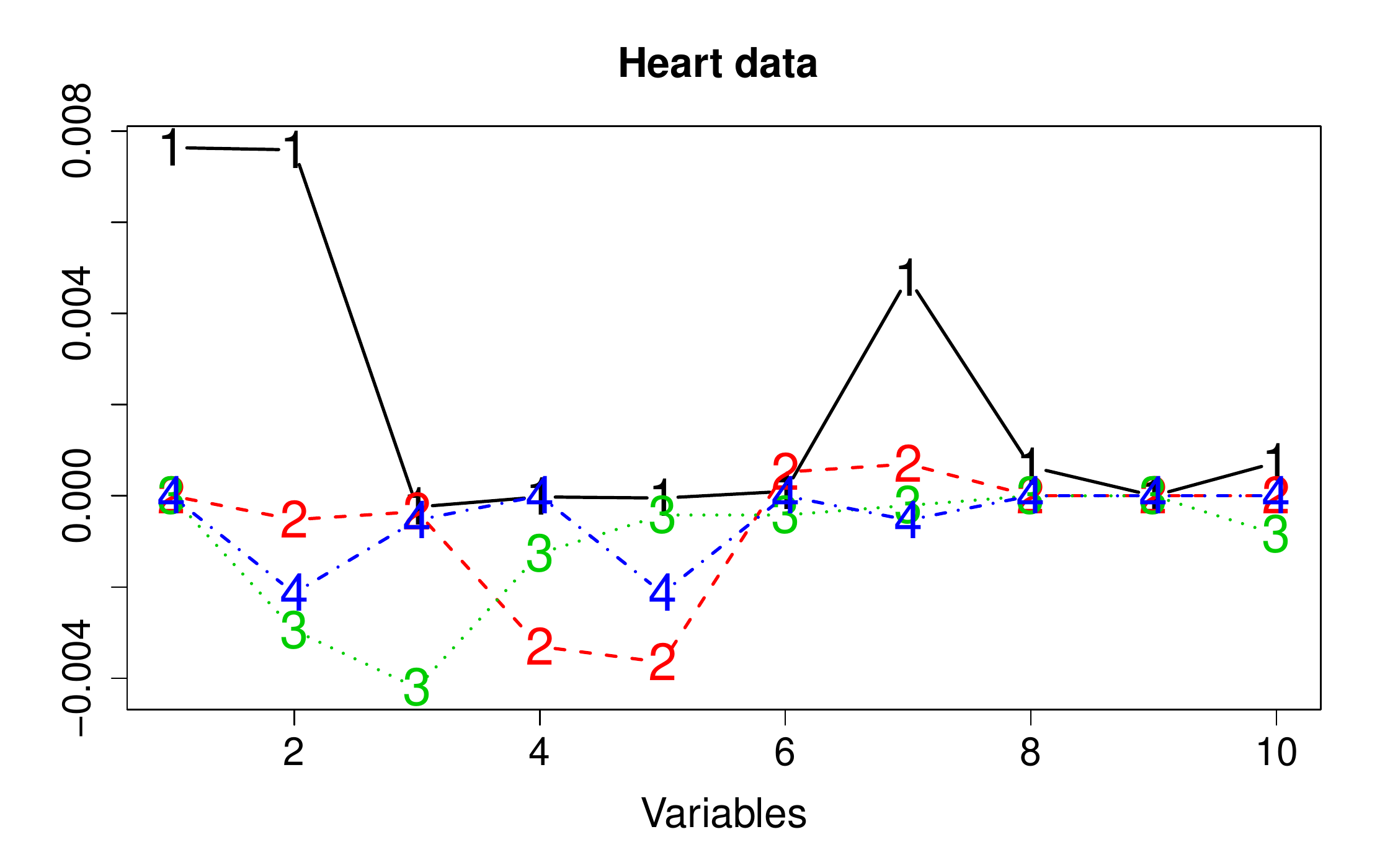}
\includegraphics[width=7cm]{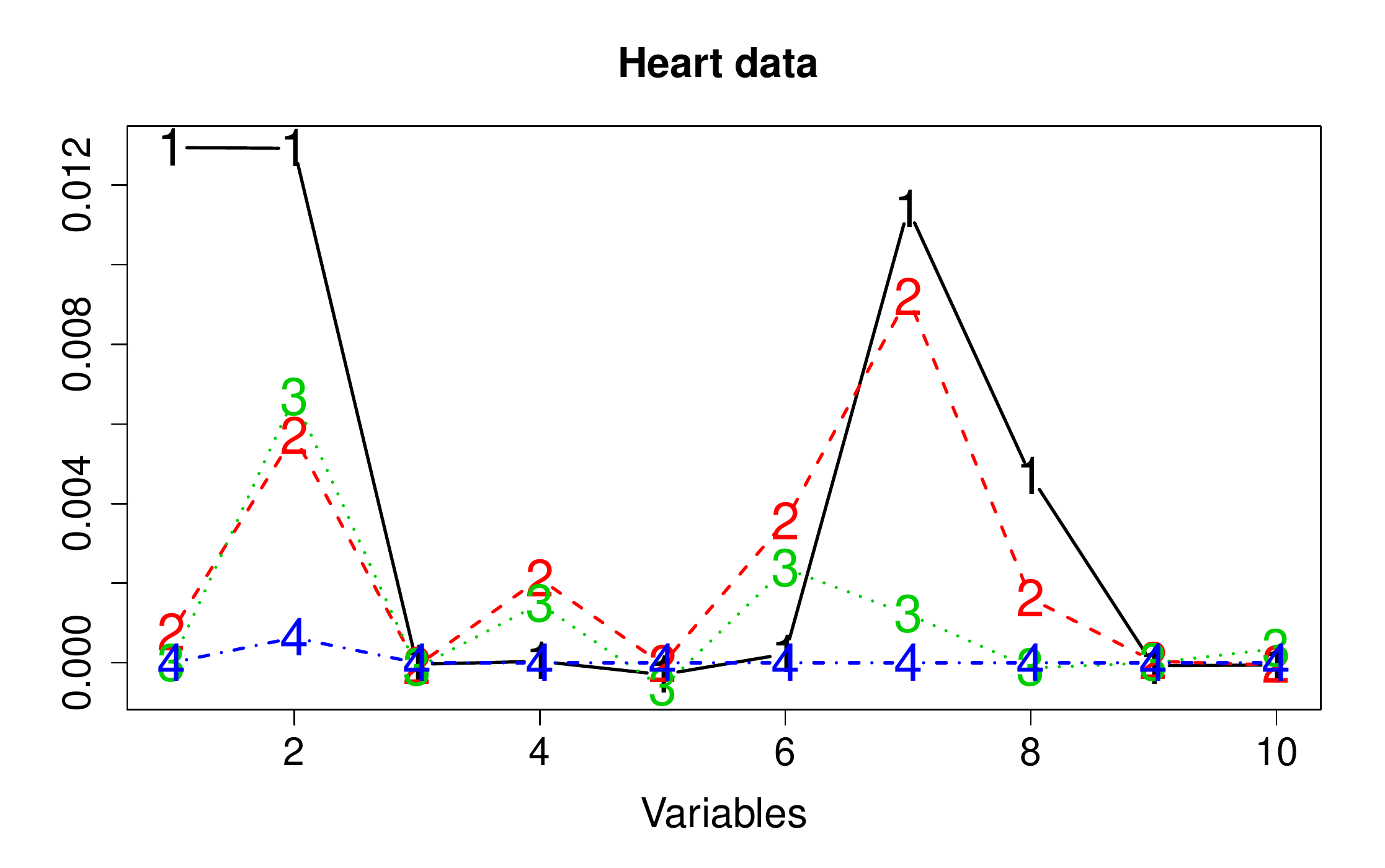}
\includegraphics[width=7cm]{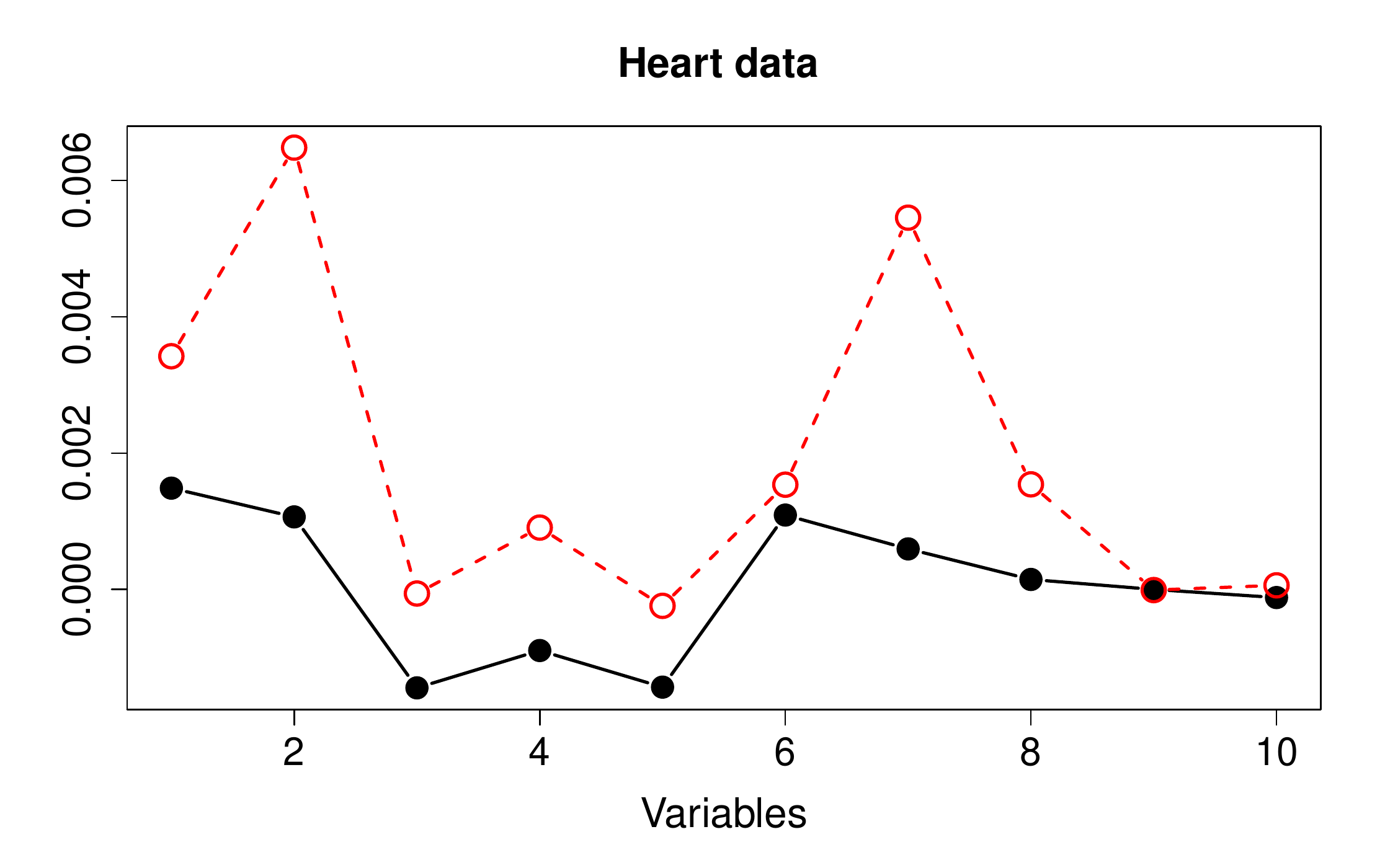}
\caption{Importance for heart data; variables 1 to 10: chest pain, oldpeak, age, trestbps, chol, thalach, exang, sex, fbs, restecg (cforest fit); left upper panel: importance for conditional splits in adjacent categories RF, right upper panel: importance for  splits in split-based RF, lower panel: averaged importance measures for splits in adjacent categories (lower curve) and split-based RF  (upper curve). }
\label{fig:heartcf}
\end{figure}

Figure \ref{fig:heartcf} shows the corresponding picture if conditional trees are used (cforest). Conditional trees avoid the bias that is found if categorical variables with varying numbers of categories and a mixture of categorical and continuous predictors are used, see, for example, \citep{strobl2007bias}. Consequently, the obtained importance measures differ from the Gini importance measures. It is seen that  variable 1,2 and 7 are very influential. In particular the importance of variable 1, which is a categorical variable, is more distinct than in Gini importance measures.

\section{Concluding Remarks}
The split  variables, which are the building blocks of ordinal models, have  been used to develop ordinal  trees and random forests. The basic concept can also be used to generate alternative parametric or nonparametric classification methods that account for the order in responses. One can, for example, use two-class linear discriminant analysis or binary models with variables selection by lasso in the case of many predictors, or use nonparametric methods as the nearest neighborhood  classifier for two classes. All of these methods can be used to model the split variables conditionally or unconditionally. In the present paper we restricted consideration to random forests since the objective was to construct score-free random forests.

Also the more recently proposed ordinal random forests are in some way inspired by parametric ordinal models but in a different way than  the split variables approach propagated here. 
The score-free random forests proposed by \citet{buri2020model} follow a quite different strategy to obtain random forests. They fit a cumulative logit model and use the likelihood contributions of the observations to obtain test statistics. The core idea is to regress the obtained partial derivatives of the log-likelihood on prognostic variables. By using the cumulative model the order of categories is used without the need for assigned scores. But it should be noted that the ``pure'' cumulative model is fitted in subpopulations without including predictors.
The ordinal forest propagated by \citet{hornung2019ordinal} also uses the cumulative logistic model. It exploits the latent continuous response variable underlying the observed ordinal response variable by explicitly using 
the widths of the adjacent intervals in the range of the  continuous response variable. These intervals are considered as corresponding
to the  classes of the ordinal response variable. 
That means, ``the ordinal response variable is treated as
a continuous variable, where the differing extents of the individual classes of the ordinal response variable are implicitly taken into account'' \citep{hornung2019ordinal}. The approach is closely related to conventional random forests for continuous outcomes but optimizes the assigned scores instead of considering them as given, and therefore  is  score-free in a certain sense.

\bibliography{literatur}
\end{document}

%% file: def.tex
\newcommand{\med}{\operatorname{med}}  %Median

\newcommand{\tr}{\operatorname{tr}}

\newcommand{\betab}{{\boldsymbol{\beta}}}

\newcommand{\gammab}{\boldsymbol{\gamma}}
\newcommand{\pib}{{\boldsymbol{\pi}}}

\newcommand{\xb}{\boldsymbol{x}}

\newcommand{\zb}{\boldsymbol{z}}

\newcommand{\blanco}[1]{}

\def\d{\displaystyle}

%% file: OrdinalRandomForests2.bbl
\begin{thebibliography}{}

\bibitem[\protect\citeauthoryear{Agresti}{Agresti}{2010}]{Agresti:2009}
Agresti, A. (2010).
\newblock {\em Analysis of Ordinal Categorical Data, 2nd Edition}.
\newblock New York: Wiley.

\bibitem[\protect\citeauthoryear{Anderson}{Anderson}{1984}]{Anderson:84}
Anderson, J.~A. (1984).
\newblock Regression and ordered categorical variables.
\newblock {\em Journal of the Royal Statistical Society\/}~{\em B 46}, 1--30.

\bibitem[\protect\citeauthoryear{Anderson and Phillips}{Anderson and
  Phillips}{1981}]{AndPhi:81}
Anderson, J.~A. and R.~R. Phillips (1981).
\newblock Regression, discrimination and measurement models for ordered
  categorical variables.
\newblock {\em Applied Statistics\/}~{\em 30}, 22--31.

\bibitem[\protect\citeauthoryear{Andrich}{Andrich}{2013}]{andrich2013expanded}
Andrich, D. (2013).
\newblock An expanded derivation of the threshold structure of the polytomous
  {R}asch model that dispels any 'threshold disorder controversy'.
\newblock {\em Educational and Psychological Measurement\/}~{\em 73\/}(1),
  78--124.

\bibitem[\protect\citeauthoryear{Archer}{Archer}{2010}]{archer2010rpartordinal}
Archer, K.~J. (2010).
\newblock rpartordinal: an {R} package for deriving a classification tree for
  predicting an ordinal response.
\newblock {\em Journal of Statistical Software\/}~{\em 34}, 7.

\bibitem[\protect\citeauthoryear{Bender and Grouven}{Bender and
  Grouven}{1998}]{BenGro:98}
Bender, R. and U.~Grouven (1998).
\newblock Using binary logistic regression models for ordinal data with
  non--proportional odds.
\newblock {\em Journal of Clinical Epidemiology\/}~{\em 51}, 809--816.

\bibitem[\protect\citeauthoryear{Brant}{Brant}{1990}]{Brant:90}
Brant, R. (1990).
\newblock Assessing proportionality in the proportional odds model for ordinal
  logistic regression.
\newblock {\em Biometrics\/}~{\em 46}, 1171--1178.

\bibitem[\protect\citeauthoryear{Breiman}{Breiman}{1996}]{Breiman:96}
Breiman, L. (1996).
\newblock Bagging predictors.
\newblock {\em Machine Learning\/}~{\em 24}, 123--140.

\bibitem[\protect\citeauthoryear{Breiman}{Breiman}{2001}]{Breiman:2001a}
Breiman, L. (2001).
\newblock Random forests.
\newblock {\em Machine Learning\/}~{\em 45}, 5--32.

\bibitem[\protect\citeauthoryear{B{\"u}hlmann, Yu, et~al.}{B{\"u}hlmann
  et~al.}{2002}]{buhlmann2002analyzing}
B{\"u}hlmann, P., B.~Yu, et~al. (2002).
\newblock Analyzing bagging.
\newblock {\em The Annals of Statistics\/}~{\em 30\/}(4), 927--961.

\bibitem[\protect\citeauthoryear{Buri and Hothorn}{Buri and
  Hothorn}{2020}]{buri2020model}
Buri, M. and T.~Hothorn (2020).
\newblock Model-based random forests for ordinal regression.
\newblock {\em The International Journal of Biostatistics\/}~{\em
  1\/}(ahead-of-print).

\bibitem[\protect\citeauthoryear{Campbell and Donner}{Campbell and
  Donner}{1989}]{CamDon:89}
Campbell, M.~K. and A.~P. Donner (1989).
\newblock Classification efficiency of multinomial logistic-regression relative
  to ordinal logistic-regression.
\newblock {\em Journal of the American Statistical Association\/}~{\em
  84\/}(406), 587--591.

\bibitem[\protect\citeauthoryear{Campbell, Donner, and Webster}{Campbell
  et~al.}{1991}]{Cam-etal:91}
Campbell, M.~K., A.~P. Donner, and K.~M. Webster (1991).
\newblock Are ordinal models useful for classification?
\newblock {\em Statistics in Medicine\/}~{\em 10}, 383--394.

\bibitem[\protect\citeauthoryear{Chu and Keerthi}{Chu and
  Keerthi}{2007}]{chu2007support}
Chu, W. and S.~S. Keerthi (2007).
\newblock Support vector ordinal regression.
\newblock {\em Neural computation\/}~{\em 19\/}(3), 792--815.

\bibitem[\protect\citeauthoryear{Cox}{Cox}{1995}]{Cox:95}
Cox, C. (1995).
\newblock Location-scale cumulative odds models for ordinal data: A generalized
  non-linear model approach.
\newblock {\em Statistics in Medicine\/}~{\em 14}, 1191--1203.

\bibitem[\protect\citeauthoryear{Fernandez, Liu, and Costilla}{Fernandez
  et~al.}{2019}]{fernandez2019method}
Fernandez, D., I.~Liu, and R.~Costilla (2019).
\newblock A method for ordinal outcomes: The ordered stereotype model.
\newblock {\em International Journal of Methods in Psychiatric Research\/},
  e1801.

\bibitem[\protect\citeauthoryear{Galimberti, Soffritti, and Di~Maso}{Galimberti
  et~al.}{2012}]{galimberti2012classification}
Galimberti, G., G.~Soffritti, and M.~Di~Maso (2012).
\newblock Classification trees for ordinal responses in {R}: the rpartscore
  package.
\newblock {\em Journal of Statistical Software\/}~{\em 47}.

\bibitem[\protect\citeauthoryear{Gneiting and Raftery}{Gneiting and
  Raftery}{2007}]{Gneitingetal:2007}
Gneiting, T. and A.~Raftery ({2007}).
\newblock {Strictly proper scoring rules, prediction, and estimation}.
\newblock {\em Journal of the American Statistical Association\/}~{\em
  {102}\/}({477}), 359--376.

\bibitem[\protect\citeauthoryear{Goodman}{Goodman}{1981a}]{Goodman:81a}
Goodman, L.~A. (1981a).
\newblock Association models and canonical correlation in the analysis of
  cross-classification having ordered categories.
\newblock {\em Journal of the American Statistical Association\/}~{\em 76},
  320--334.

\bibitem[\protect\citeauthoryear{Goodman}{Goodman}{1981b}]{Goodman:81b}
Goodman, L.~A. (1981b).
\newblock Association models and the bivariate normal for contingency tables
  with ordered categories.
\newblock {\em Biometrika\/}~{\em 68}, 347--355.

\bibitem[\protect\citeauthoryear{Greenland}{Greenland}{1994}]{Greenland:94}
Greenland, S. (1994).
\newblock Alternative models for ordinal logistic regression.
\newblock {\em Statistics in Medicine\/}~{\em 13}, 1665--1677.

\bibitem[\protect\citeauthoryear{Gregorutti, Michel, and
  Saint-Pierre}{Gregorutti et~al.}{2017}]{gregorutti2017correlation}
Gregorutti, B., B.~Michel, and P.~Saint-Pierre (2017).
\newblock Correlation and variable importance in random forests.
\newblock {\em Statistics and Computing\/}~{\em 27\/}(3), 659--678.

\bibitem[\protect\citeauthoryear{Hapfelmeier, Hothorn, Ulm, and
  Strobl}{Hapfelmeier et~al.}{2014}]{hapfelmeier2014new}
Hapfelmeier, A., T.~Hothorn, K.~Ulm, and C.~Strobl (2014).
\newblock A new variable importance measure for random forests with missing
  data.
\newblock {\em Statistics and Computing\/}~{\em 24\/}(1), 21--34.

\bibitem[\protect\citeauthoryear{Hornung}{Hornung}{2020}]{hornung2019ordinal}
Hornung, R. (2020).
\newblock Ordinal forests.
\newblock {\em Journal of Classification\/}~{\em 37}, 4--17.

\bibitem[\protect\citeauthoryear{Hothorn, Hornik, and Zeileis}{Hothorn
  et~al.}{2006}]{Hotetal:2006}
Hothorn, T., K.~Hornik, and A.~Zeileis ({2006}).
\newblock {Unbiased recursive partitioning: A conditional inference framework}.
\newblock {\em {Journal of Computational and Graphical Statistics}\/}~{\em
  {15}}, {651--674}.

\bibitem[\protect\citeauthoryear{Hothorn and Zeileis}{Hothorn and
  Zeileis}{2015}]{hothorn2015partykit}
Hothorn, T. and A.~Zeileis (2015).
\newblock partykit: A modular toolkit for recursive partytioning in r.
\newblock {\em The Journal of Machine Learning Research\/}~{\em 16\/}(1),
  3905--3909.

\bibitem[\protect\citeauthoryear{Iannario, Piccolo, and Simone}{Iannario
  et~al.}{2020}]{iannario2018cub}
Iannario, M., D.~Piccolo, and R.~Simone (2020).
\newblock {CUB}: a class of mixture models for ordinal data. {R} package
  version 1.1.4, http://cran.r-project.org/package=cub.

\bibitem[\protect\citeauthoryear{Janitza, Tutz, and Boulesteix}{Janitza
  et~al.}{2016}]{janitza2016random}
Janitza, S., G.~Tutz, and A.-L. Boulesteix (2016).
\newblock Random forest for ordinal responses: prediction and variable
  selection.
\newblock {\em Computational Statistics \& Data Analysis\/}~{\em 96}, 57--73.

\bibitem[\protect\citeauthoryear{Kateri}{Kateri}{2014}]{kateri2014contingency}
Kateri, M. (2014).
\newblock {\em Contingency table analysis}.
\newblock Springer.

\bibitem[\protect\citeauthoryear{Kim}{Kim}{2003}]{kim2003assessing}
Kim, J.-H. (2003).
\newblock Assessing practical significance of the proportional odds assumption.
\newblock {\em Statistics \& probability letters\/}~{\em 65\/}(3), 233--239.

\bibitem[\protect\citeauthoryear{Liaw, Wiener, Breiman, and Cutler}{Liaw
  et~al.}{2015}]{liaw2015package}
Liaw, A., M.~Wiener, L.~Breiman, and A.~Cutler (2015).
\newblock Package randomforest.

\bibitem[\protect\citeauthoryear{Liu, Mukherjee, Suesse, Sparrow, and Park}{Liu
  et~al.}{2009}]{liu2009graphical}
Liu, I., B.~Mukherjee, T.~Suesse, D.~Sparrow, and S.~K. Park (2009).
\newblock Graphical diagnostics to check model misspecification for the
  proportional odds regression model.
\newblock {\em Statistics in medicine\/}~{\em 28\/}(3), 412--429.

\bibitem[\protect\citeauthoryear{Masters}{Masters}{1982}]{Masters:82}
Masters, G.~N. (1982).
\newblock A {R}asch model for partial credit scoring.
\newblock {\em Psychometrika\/}~{\em 47}, 149--174.

\bibitem[\protect\citeauthoryear{Masters and Wright}{Masters and
  Wright}{1984}]{MasWri:84}
Masters, G.~N. and B.~Wright (1984).
\newblock The essential process in a family of measurement models.
\newblock {\em Psychometrika\/}~{\em 49}, 529--544.

\bibitem[\protect\citeauthoryear{McCullagh}{McCullagh}{1980}]{McCullagh:80}
McCullagh, P. (1980).
\newblock Regression model for ordinal data (with discussion).
\newblock {\em Journal of the Royal Statistical Society\/}~{\em B 42},
  109--127.

\bibitem[\protect\citeauthoryear{Muraki}{Muraki}{1997}]{muraki1997generalized}
Muraki, E. (1997).
\newblock A generalized partial credit model.
\newblock {\em Handbook of modern item response theory\/}, 153--164.

\bibitem[\protect\citeauthoryear{Peterson and Harrell}{Peterson and
  Harrell}{1990}]{PetHar:90}
Peterson, B. and F.~E. Harrell (1990).
\newblock Partial proportional odds models for ordinal response variables.
\newblock {\em Applied Statistics\/}~{\em 39}, 205--217.

\bibitem[\protect\citeauthoryear{Rattinger, Ro{\ss}teutscher, Schmitt-Beck,
  We{\ss}els, and Wolf}{Rattinger et~al.}{2014}]{GLES}
Rattinger, H., S.~Ro{\ss}teutscher, R.~Schmitt-Beck, B.~We{\ss}els, and C.~Wolf
  (2014).
\newblock Pre-election cross section {(GLES 2013)}.
\newblock {\em GESIS Data Archive, Cologne\/}~{\em ZA5700 Data file Version
  2.0.0}.

\bibitem[\protect\citeauthoryear{Rudolfer, Watson, and Lesaffre}{Rudolfer
  et~al.}{1995}]{Rud-etal:95}
Rudolfer, S.~M., P.~C. Watson, and E.~Lesaffre (1995).
\newblock Are ordinal models useful for classification? a revised analysis.
\newblock {\em Journal of Statistical Computation Simulation\/}~{\em 52\/}(2),
  105--132.

\bibitem[\protect\citeauthoryear{Strobl, Boulesteix, Kneib, Augustin, and
  Zeileis}{Strobl et~al.}{2008}]{strobl2008conditional}
Strobl, C., A.-L. Boulesteix, T.~Kneib, T.~Augustin, and A.~Zeileis (2008).
\newblock Conditional variable importance for random forests.
\newblock {\em BMC bioinformatics\/}~{\em 9\/}(1), 307.

\bibitem[\protect\citeauthoryear{Strobl, Boulesteix, Zeileis, and
  Hothorn}{Strobl et~al.}{2007}]{strobl2007bias}
Strobl, C., A.-L. Boulesteix, A.~Zeileis, and T.~Hothorn (2007).
\newblock Bias in random forest variable importance measures: Illustrations,
  sources and a solution.
\newblock {\em BMC bioinformatics\/}~{\em 8\/}(1), 25.

\bibitem[\protect\citeauthoryear{Tutz}{Tutz}{2012}]{TutzBook2011}
Tutz, G. (2012).
\newblock {\em {Regression for Categorical Data}}.
\newblock Cambridge University Press.

\bibitem[\protect\citeauthoryear{Tutz}{Tutz}{2021}]{TuHetWire21}
Tutz, G. (2021).
\newblock Ordinal regression: a review and a taxonomy of models.
\newblock {\em WIRES Computational Statistics, to appear\/}.

\end{thebibliography}
